\documentclass[twocolumn,showpacs,amsmath,amssymb,superscriptaddress]{revtex4-1}
\usepackage{dcolumn}
\usepackage{color}
\usepackage{graphicx}
\usepackage{amsmath}
\usepackage{multirow}
\usepackage{bm}
\usepackage{url}

\begin{document}

\title{Description of odd-mass nuclei within the interacting boson-fermion model based on the Gogny
energy density functional}

\author{K.~Nomura}
\affiliation{Physics Department, Faculty of Science, University of
Zagreb, HR-10000 Zagreb, Croatia}
\affiliation{Center for Computational
Sciences, University of Tsukuba, Tsukuba 305-8577, Japan}

\author{R.~Rodr\'iguez-Guzm\'an}
\affiliation{Physics Department, Kuwait University, 13060 Kuwait, Kuwait}

\author{L.~M.~Robledo}
\affiliation{Departamento de F\'\i sica Te\'orica, Universidad
Aut\'onoma de Madrid, E-28049 Madrid, Spain}

\date{\today}

\begin{abstract}
Spectroscopic properties of odd-mass nuclei are studied within the 
framework of the interacting boson-fermion model (IBFM) with parameters 
based on the  Hartree-Fock-Bogoliubov (HFB) approximation. The 
parametrization D1M of the Gogny energy density functional (EDF) has 
been used at the mean-field level to obtain the deformation energy 
surfaces for the considered nuclei in terms of the quadrupole 
deformations ($\beta,\gamma$). In addition to the energy surfaces, both  
single particle energies and occupation probabilities have been used as 
a microscopic input for building the IBFM Hamiltonian. Only three 
strength parameters for the particle-boson-core coupling are fitted to 
experimental spectra. The IBFM Hamiltonian is then used to compute the 
energy spectra and electromagnetic transition rates for selected  
odd-mass Eu and Sm nuclei as well as for $^{195}$Pt and $^{195}$Au. A 
reasonable agreement with the available experimental data is obtained 
for the considered odd-mass nuclei.
\end{abstract}


\maketitle


\section{Introduction}


The study of the nuclei with odd number of protons $Z$ and/or neutrons 
$N$ has been always a recurrent theme of great interest in low-energy 
nuclear physics. While a wealth of new spectroscopic data have been 
produced experimentally for odd nuclei, especially in medium- and 
heavy-mass regions, a microscopic description of  odd nuclei 
represents one of the most challenging problems on the theoretical side. One 
of the reasons is the treatment of pairing: this type of correlation 
couples protons and neutrons in even-even nuclei to form Cooper pairs 
similar to the ones of the phenomenon of superfluidity in condensed 
matter physics. The  presence of that Cooper pairs influence in an important 
way nuclear dynamics as to determine basic parameters of vibrational 
and rotational spectra \cite{BM}. The treatment of pairing correlations 
in those cases relies on the Hartree-Fock-Bogoliubov (HFB) mean field 
theory characteristic of even-even nuclei. On the other hand, the 
description of odd-$Z$ and/or $N$ nuclei requires to describe  at the 
same time the Cooper pairs and the unpaired nucleon, that is boson like 
(Cooper pairs) and fermionic (unpaired nucleon) degrees of freedom. The 
treatment of both degrees of freedom requires to extend the HFB theory to include blocking with 
the subsequent complications  arising by the explicit 
breaking of time reversal invariance \cite{BM,bohr1953}. Another 
unwanted consequence is the  need to explicitly treat both the 
collective and single-particle degrees of freedom on the same footing.

The above difficulties are not a problem for the shell model (SM) 
\cite{caurier2005}, but the exceedingly large size of shell model 
configuration spaces in medium mass and heavy nuclei and/or open shell 
nuclei make it impractical for the purpose of describing odd nuclei in 
those cases. 

On the other hand, the theoretical approaches based on the energy
density functional (EDF) method \cite{bender2003} allow for a global 
description of many low energy  properties of nuclei all over
the nuclide chart, including medium-mass and heavy ones. 
Although not as common as in the even-even case, a number of 
calculations have been made within the EDF framework at the mean-field 
level for odd-mass systems (see 
\cite{PhysRevC.86.064313,rayner10odd-1,rayner10odd-2,rayner10odd-3} for 
some recent examples with the Gogny force). In the EDF framework, a 
proper description of excited states requires the inclusion of 
dynamical correlations associated with the restoration of broken 
symmetries and fluctuations via the symmetry-projected configuration 
mixing calculation within the generator coordinate method (GCM). In 
Ref.~\cite{bally2014}, the GCM framework has been extended to the 
odd-mass systems by explicitly taking into account the breaking of 
time-reversal symmetry. Nevertheless, the practical applications of this 
approach to medium-heavy and heavy nuclei are computationally 
demanding, and  so far calculations have been limited to very light-mass 
systems \cite{bally2014,Borrajo2016}.

Given the difficulties encountered with the two major theoretical
approaches to nuclear structure mentioned earlier, i.e., SM and EDF-GCM
frameworks, it is worth to consider alternative theoretical approaches
to odd nuclei. 
Among them we can mention the various extensions of the
particle-vibration coupling scheme
\cite{BM,litvinova2011,mizuyama2012,niu2015,degregorio2016,degregorio2017} 
and also algebraic based approaches \cite{scholten1985,IBFM,IBFM-Book,iachello2011,petrellis2011}, that 
provide a computationally more economic as well as straightforward 
description of the odd nuclei in all mass regions from light to heavy 
nuclei. Along this direction, one of the present authors has recently 
developed a method \cite{nomura2016odd} to calculate the spectroscopic 
properties of odd-mass nuclei, which is based on the EDF framework 
combined with the particle-boson-core coupling scheme, i.e., the 
interacting boson-fermion model \cite{IBFM}.  In this proposal 
\cite{nomura2016odd}, the energy surface of an even-even nucleus given 
as a function of the shape parameters describing quadrupole deformation 
$(\beta,\gamma)$ along with  single-particle energies and occupation 
probabilities of the odd nucleons computed within the self-consistent 
mean-field method (based on the relativistic EDF with the 
parametrization DD-PC1 \cite{DDPC1}) are used as a microscopic input to 
determine the parameters of an IBFM Hamiltonian. Only the strength 
parameters of the boson-fermion coupling term in the IBFM Hamiltonian 
have been determined so as to fit the selected experimental data. The 
validity of the method has been confirmed  in the axially-deformed 
odd-mass Eu isotopes \cite{nomura2016odd} and further applied to study  
shape phase transitions between nearly spherical and axially deformed 
shapes in the odd-mass Eu and Sm isotopes \cite{nomura2016qpt}, and
between nearly spherical and $\gamma$-soft shapes in the mass $A\approx
130$ region \cite{nomura2017gsoft}.

In this work, we apply the method of Ref.~\cite{nomura2016odd} to 
describe spectroscopic properties of selected odd-mass nuclei based on 
the Gogny EDF. Specifically, we consider the odd-mass isotopes 
$^{149-155}$Eu and $^{149-155}$Sm as well as the $^{195}$Pt and the 
$^{195}$Au nuclei. Their associated  even-even core nuclei, 
$^{148-154}$Sm and $^{194}$Pt, are good examples of axially-deformed 
and $\gamma$-soft nuclei, respectively. By studying the same isotopes 
($^{149-155}$Eu and $^{149-155}$Sm) as in 
Refs.~\cite{nomura2016odd,nomura2016qpt}, we demonstrate that the IBFM 
Hamiltonian based on the Gogny EDF describes the low-lying states in 
odd-mass nuclei at the same level of accuracy as in the earlier studies
of \cite{nomura2016odd,nomura2016qpt}  based on the relativistic EDF. 
The Gogny EDF is a successful member of the class of
non-relativistic EDF. It has been used in many nuclear structure and reaction
theory calculations all over the nuclide chart with great success.
Its accuracy in describing experimental data is at the level of the 
one obtained with performing modern relativistic EDF.
By comparing the results within the two major classes of EDF,
i.e., non-relativistic and relativistic EDFs, we demonstrate the
validity of the mapping procedure of Ref.~\cite{nomura2016odd}. 
Furthermore, the addition of the $\gamma$-soft nuclei, $^{195}$Pt and
$^{195}$Au, which have not been included in
Refs.~\cite{nomura2016odd,nomura2016qpt}, further confirms the
robustness of the procedure. 

We employ the parametrization D1M \cite{D1M} of the Gogny EDF. A number 
of previous studies have demonstrated that the D1M set, apart from 
being much more efficient in the description of binding energies than 
the more traditional and extensively tested D1S parametrization 
\cite{D1S}, it keeps the same predictive power as D1S in the 
description of other nuclear properties like excitation energies or 
transition strengths. Nevertheless, we have also carried out part of 
the calculations with D1S and confirmed the striking similarities 
between both sets of  results. As a consequence, throughout the paper 
we will only discuss the results obtained with the parametrization D1M.

In Sec.~\ref{sec:model} we give a brief account of the method used to 
describe the considered odd-mass nuclei, and then present the 
parameters for the boson-core Hamiltonian, single-particle energies and 
occupation probabilities for odd particle, and the fitted strength 
parameters for the boson-fermion coupling interaction. In 
Sec.~\ref{sec:pes}, we present the results for the even-even core 
nuclei, including the Gogny-D1M and mapped energy surfaces, and the 
calculated low-energy excitation spectra in comparison to the 
experimental data. In Sec.~\ref{sec:eusm}, the results for the 
spectroscopic calculation for the odd-mass Eu and Sm isotopes, 
including evolution of energy levels, $B(E2)$, and spectroscopic quadrupole
and magnetic moments, are discussed. In 
Sec.~\ref{sec:ptau} the energy spectra and decay patterns in the 
$\gamma$-soft nuclei $^{195}$Pt and $^{195}$Au are discussed and 
compared with the available experimental data. Finally, 
Sec.~\ref{sec:summary} is devoted to the summary of the paper and to 
discuss future  perspectives.


\section{Description of the model\label{sec:model}}


In this section, we briefly describe the theoretical  framework 
proposed in Ref \cite{nomura2016odd} and used in this study. We also 
discuss the parameters of the IBFM Hamiltonian employed in the 
calculations. For more details on the philosophy of the model as well 
as its main assumptions the reader is referred to 
Ref.~\cite{nomura2016odd} for a thorough discussion.

\subsection{Construction of IBFM Hamiltonian}

The IBFM Hamiltonian, used to describe the studied odd-mass  nuclei, 
consists of three terms, namely, the even-even boson core or 
Interacting Boson Model (IBM) Hamiltonian $\hat H_B$, the 
single-particle Hamiltonian for unpaired fermions $\hat H_F$ and the 
boson-fermion coupling term $\hat H_{BF}$
\begin{equation}
\label{eq:ham}
 \hat H=\hat H_B + \hat H_F + \hat H_{BF}. 
\end{equation}

The building blocks of the IBM are the traditional $s$ and $d$ bosons, which
represent the collective pairs of valence nucleons \cite{OAI} coupled
to angular momentum $J^\pi=0^+$ and $2^+$, respectively. 
The number of bosons $N_B$ and fermions $N_F$ are assumed to be
conserved separately. Note also that we use the traditional version of the IBM
where there is no distinction between neutron and 
proton bosons. Finally, we assume that as we only consider odd-mass nuclei, the 
number of fermions equals one $N_F=1$. The IBM Hamiltonian $\hat H_B$ reads
\begin{equation}
\label{eq:ibm}
 \hat H_B = \epsilon_d\hat n_d + \kappa\hat Q_B\cdot\hat Q_B +
  \kappa^{\prime}\hat L\cdot\hat L, 
\end{equation}
given in terms of the $d$-boson number operator $\hat n_d=d^{\dagger}\cdot\tilde d$,
the quadrupole operator $\hat Q_B=s^{\dagger}\tilde d+d^{\dagger}\tilde s +
\chi[d^{\dagger}\times\tilde d]^{(2)}$, and the angular momentum
operator $\hat L=\sqrt{10}[d^{\dagger}\times\tilde d]^{(1)}$. The remaining 
quantities $\epsilon_d$, $\kappa$, $\kappa^{\prime}$ and $\chi$ 
are parameters of the Hamiltonian $\hat H_{B}$.

On the other hand, the single-fermion Hamiltonian takes the form
\begin{equation}
\hat
H_F=\sum_{j}\epsilon_j[a^{\dagger}_j\times\tilde a_j]^{(0)} 
\end{equation} 
where $a^{\dagger}_j$ and $a_j$ are the fermion creation and annihilation
operators while $\epsilon_j$ stands for the single-particle energy of the 
orbital $j$.

For the boson-fermion coupling Hamiltonian $\hat H_{BF}$ we have employed the  
simplest possible form, as suggested in Refs~\cite{IBFM,scholten1985}
\begin{eqnarray}
\label{eq:bf}
 \hat H_{BF}
=&&\sum_{jj^{\prime}}\Gamma_{jj^{\prime}}\hat
  Q_B\cdot[a^{\dagger}_j\times\tilde a_{j^{\prime}}]^{(2)} \nonumber \\
&&+\sum_{jj^{\prime}j^{\prime\prime}}\Lambda_{jj^{\prime}}^{j^{\prime\prime}}
:[[d^{\dagger}\times\tilde a_{j}]^{(j^{\prime\prime})}
\times
[a^{\dagger}_{j^{\prime}}\times\tilde d]^{(j^{\prime\prime})}]^{(0)}:
\nonumber \\
&&+\sum_j A_j[a^{\dagger}\times\tilde a_{j}]^{(0)}\hat n_d, 
\end{eqnarray}
where the first, second and third terms are referred to as the quadrupole, 
exchange, and monopole interactions, respectively. The strength parameters $\Gamma_{jj^{\prime}}$,
$\Lambda_{jj^{\prime}}^{j^{\prime\prime}}$ and $A_j$ can be 
expressed, using the generalized seniority scheme, in the following
$j$-dependent forms \cite{scholten1985}
\begin{eqnarray}
\label{eq:dynamical}
&&\Gamma_{jj^{\prime}}=\Gamma_0\gamma_{jj^{\prime}} \\
\label{eq:exchange}
&&\Lambda_{jj^{\prime}}^{j^{\prime\prime}}=-2\Lambda_0\sqrt{\frac{5}{2j^{\prime\prime}+1}}\beta
_{jj^{\prime\prime}}\beta_{j^{\prime}j^{\prime\prime}} \\
\label{eq:monopole}
&&A_j=-A_0\sqrt{2j+1}
\end{eqnarray}
where
$\gamma_{jj^{\prime}}=(u_ju_{j^{\prime}}-v_jv_{j^{\prime}})Q_{jj^{\prime}}$
and 
$\beta_{jj^{\prime}}=(u_jv_{j^{\prime}}+v_ju_{j^{\prime}})Q_{jj^{\prime}}$,
with the matrix element of the quadrupole operator in the single-particle
basis $Q_{jj^{\prime}}=\langle j||Y^{(2)}||j^{\prime}\rangle$.
Both $u_j$ and $v_j$ represent  the occupation probabilities for the
orbital $j$ and  satisfy the well known relation $u_j^2+v_j^2=1$. 
Furthermore, $\Gamma_0$, $\Lambda_0$ and $A_0$ denote the 
corresponding strength parameters.
Note that an exhaustive presentation of the physical contents of the
formulas in Eqs.~(\ref{eq:bf})-(\ref{eq:monopole}) as
well as the discussion of relevant applications to  odd-A nuclei, has
already been considered in Ref.~\cite{scholten1985}.

The first step to build the IBFM Hamiltonian $\hat H$ in 
Eq.~(\ref{eq:ham}) is to determine the IBM Hamiltonian $\hat H_B$ by 
using the fermion-to-boson mapping procedure developed in 
Refs.~\cite{nomura2008,nomura2010,nomura2011rot}. Here, the fermion 
($\beta\gamma$)-deformation energy surface, obtained within the 
constrained Gogny-D1M HFB framework, is mapped onto the expectation 
value of $\hat H_B$ in the boson condensate state \cite{ginocchio1980}. 
This procedure completely determines the parameters $\epsilon_d$, 
$\kappa$ and $\chi$. The strength parameter $\kappa^{\prime}$ for the 
$\hat L\cdot\hat L$ term is obtained separately by equating the 
cranking moment of inertia, calculated in the boson coherent state at 
the energy minimum, to the corresponding Thouless-Valatin moment of 
inertia computed within the cranked HFB approach \cite{nomura2011rot}. 
It has been shown in Ref.~\cite{nomura2011rot} that the $\hat 
L\cdot\hat L$ term is only relevant in axially-deformed systems and, 
for that reason, we do not included it in the calculation for the 
$\gamma$-soft nucleus $^{194}$Pt. For a more detailed account of 
constrained Gogny-HFB calculations the reader is referred, for example, 
to Refs.~\cite{robledo2008,rayner2010pt}. Details of the 
fermion-to-boson mapping procedure are given in  
Refs.~\cite{nomura2008,nomura2010}. The parameters derived for the 
isotopes  $^{148-154}$Sm and $^{194}$Pt can be found in 
Table~\ref{tab:paraB}. 

\begin{table}[hb!]
\caption{\label{tab:paraB} Parameters ($\epsilon_d$, $\kappa$, 
$\kappa^{\prime}$ and $\chi$) of the boson Hamiltonian $\hat H_B$. 
All entries, except the dimensionless parameter $\chi$, are in MeV.}
\begin{center}
\begin{tabular*}{\columnwidth}{p{1.6cm}p{1.6cm}p{1.6cm}p{1.6cm}p{1.6cm}}
\hline\hline
\textrm{} &
\textrm{$\epsilon_{d}$}&
\textrm{$\kappa$}&
\textrm{$\kappa^{\prime}$}&
\textrm{$\chi$} \\
\hline
$^{148}$Sm & 1.185 & -0.079 & -0.027 & -0.44 \\
$^{150}$Sm & 0.615 & -0.074 & -0.014 & -0.50 \\
$^{152}$Sm & 0.336 & -0.074 & -0.018 & -0.50 \\
$^{154}$Sm & 0.106 & -0.074 & -0.018 & -0.50 \\
$^{194}$Pt & 0.011 & -0.098 & - &  0.10 \\
\hline\hline
\end{tabular*}
\end{center}
\end{table}

For the fermion valence space, we have included all the spherical 
single-particle orbitals in the proton major shell $Z=50-82$ (i.e., 
$3s_{1/2}$, $2d_{3/2}$, $2d_{5/2}$ and $1g_{7/2}$ for positive-parity 
states and $1h_{11/2}$ for the intruder negative-parity states) for the 
odd-mass systems $^{149-155}$Eu and $^{195}$Au. 

On the other hand, in the case of $^{149-155}$Sm and $^{195}$Pt, with 
the Fermi level lying in the neutron major shell $N=82-126$, we have 
considered the  positive parity intruder orbital $1i_{13/2}$ and the 
negative parity orbitals $3p_{1/2}$, $3p_{3/2}$, $2f_{5/2}$, $2f_{7/2}$ 
and $1h_{9/2}$.
 
Following the prescription of Ref.~\cite{nomura2016odd}, the 
single-particle energies $\epsilon_j$ and the occupation probabilities 
$v_j^2$ are obtained from self-consistent  Gogny-D1M HFB calculations 
at the spherical configuration. In those calculations, for a given 
odd-mass nucleus with the odd neutron (proton) number $N_{0}$ 
($Z_{0}$), the standard even number parity constrained Gogny-HFB 
approach (i.e., without blocking) has been used but using $N_{0}$ 
($Z_{0}$) for the neutron (proton) number constraint. The 
single-particle energies and occupation probabilities obtained for the 
considered odd-$A$ nuclei, are shown in Tables~\ref{tab:spe} and 
\ref{tab:vv}, respectively. 

\begin{table}[hb!]
\caption{\label{tab:spe} Spherical single-particle energies (in MeV) 
 resulting from  Gogny-D1M HFB calculations  for the considered 
 odd-mass nuclei. For details, see the main text.}
\begin{center}
\begin{tabular*}{\columnwidth}{p{1.33cm}p{1.33cm}p{1.33cm}p{1.33cm}p{1.33cm}p{1.33cm}}
\hline\hline
\textrm{} &
\textrm{$3s_{1/2}$}&
\textrm{$2d_{3/2}$}&
\textrm{$2d_{5/2}$}&
\textrm{$1g_{7/2}$}&
\textrm{$1h_{11/2}$} \\
\hline
$^{149}$Eu & 3.365 & 3.076 & 0.868 & 0.000 & 3.512 \\
$^{151}$Eu & 3.378 & 3.063 & 0.850 & 0.000 & 3.544 \\
$^{153}$Eu & 3.425 & 3.078 & 0.876 & 0.000 & 3.593 \\
$^{155}$Eu & 3.494 & 3.114 & 0.936 & 0.000 & 3.653 \\
$^{195}$Au & 0.000 & 0.907 & 2.624 & 5.164 & 0.840 \\
\end{tabular*}
\begin{tabular*}{\columnwidth}{p{1.14cm}p{1.14cm}p{1.14cm}p{1.14cm}p{1.14cm}p{1.14cm}p{1.14cm}}
\hline
\textrm{} &
\textrm{$3p_{1/2}$}&
\textrm{$3p_{3/2}$}&
\textrm{$2f_{5/2}$}&
\textrm{$2f_{7/2}$}&
\textrm{$1h_{9/2}$}&
\textrm{$1i_{13/2}$} \\
\hline
$^{149}$Sm & 3.528 & 2.607 & 3.049 & 0.000 & 1.191 & 3.310 \\
$^{151}$Sm & 3.491 & 2.573 & 3.052 & 0.000 & 1.141 & 3.268 \\
$^{153}$Sm & 3.458 & 2.544 & 3.041 & 0.000 & 1.076 & 3.214 \\
$^{155}$Sm & 3.430 & 2.521 & 3.021 & 0.000 & 1.005 & 3.154 \\
$^{195}$Pt & 0.000 & 0.927 & 1.014 & 3.816 & 4.273 & 1.495 \\
\hline\hline
\end{tabular*}
\end{center}
\end{table}

\begin{table}[hb!]
\caption{\label{tab:vv} Occupation probabilities of the single-particle
 orbitals resulting from  Gogny-D1M HFB calculations for the
 considered odd-mass nuclei. For details, see the main text.}
\begin{center}
\begin{tabular*}{\columnwidth}{p{1.33cm}p{1.33cm}p{1.33cm}p{1.33cm}p{1.33cm}p{1.33cm}}
\hline\hline
\textrm{} &
\textrm{$3s_{1/2}$}&
\textrm{$2d_{3/2}$}&
\textrm{$2d_{5/2}$}&
\textrm{$1g_{7/2}$}&
\textrm{$1h_{11/2}$} \\
\hline
$^{149}$Eu & 0.112 & 0.158 & 0.700 & 0.843 & 0.102 \\
$^{151}$Eu & 0.110 & 0.159 & 0.705 & 0.845 & 0.099 \\
$^{153}$Eu & 0.107 & 0.159 & 0.706 & 0.851 & 0.095 \\
$^{155}$Eu & 0.104 & 0.159 & 0.703 & 0.858 & 0.092 \\
$^{195}$Au & 0.617 & 0.870 & 0.968 & 0.989 & 0.864 \\
\end{tabular*}
\begin{tabular*}{\columnwidth}{p{1.14cm}p{1.14cm}p{1.14cm}p{1.14cm}p{1.14cm}p{1.14cm}p{1.14cm}}
\hline
\textrm{} &
\textrm{$3p_{1/2}$}&
\textrm{$3p_{3/2}$}&
\textrm{$2f_{5/2}$}&
\textrm{$2f_{7/2}$}&
\textrm{$1h_{9/2}$}&
\textrm{$1i_{13/2}$}\\
\hline
$^{149}$Sm & 0.013 & 0.023 & 0.027 & 0.413 & 0.126 & 0.022 \\
$^{151}$Sm & 0.020 & 0.036 & 0.039 & 0.531 & 0.202 & 0.034 \\
$^{153}$Sm & 0.028 & 0.053 & 0.053 & 0.623 & 0.291 & 0.049 \\
$^{155}$Sm & 0.038 & 0.075 & 0.069 & 0.693 & 0.387 & 0.067 \\
$^{195}$Pt & 0.303 & 0.603 & 0.634 & 0.956 & 0.961 & 0.763 \\
\hline\hline
\end{tabular*}
\end{center}
\end{table}

The coupling constants of the boson-fermion interaction term $\hat 
H_{BF}$ ($\Gamma_0$, $\Lambda_0$ and $A_0$) are the only 
free parameters in our study. They are fitted, for each 
nucleus, to reproduce the lowest few experimental energy levels, 
separately for positive- and negative-parity states 
\cite{nomura2016odd}. We show in Table~\ref{tab:paraBF} the fitted 
strength parameters for the positive- ($\Gamma_0^+$, $\Lambda_0^+$ and 
$A_0^+$) and negative-parity ($\Gamma_0^-$, $\Lambda_0^-$ and $A_0^-$) 
states.

\begin{table}[hb!]
\caption{\label{tab:paraBF} Fitted parameters of the boson-fermion Hamiltonian $\hat H_{BF}$
 ($\Gamma_0^\pm$, $\Lambda_0^\pm$ and $A_0^\pm$). All entries are in MeV. }
\begin{center}
\begin{tabular*}{\columnwidth}{p{1.14cm}p{1.14cm}p{1.14cm}p{1.14cm}p{1.14cm}p{1.14cm}p{1.14cm}}
\hline\hline
\textrm{} &
\textrm{$\Gamma_0^+$}&
\textrm{$\Lambda_0^+$}&
\textrm{$A_0^+$} &
\textrm{$\Gamma_0^-$}&
\textrm{$\Lambda_0^-$}&
\textrm{$A_0^-$} \\
\hline
$^{149}$Eu & 0.05 & 2.5 & -0.13 & 0.3 & 3.5 & -0.14 \\
$^{151}$Eu & 0.06 & 1.0 & 0.0 & 0.6 & 6.5 & -0.06 \\
$^{153}$Eu & 0.17 & 7.0 & -0.65 & 0.6 & 9.0 & -0.30 \\
$^{155}$Eu & 0.19 & 4.5 & -0.44 & 0.5 & 8.0 & -0.30  \\
$^{149}$Sm & 0.2 & 36.0 & -0.25 & 0.2 & 1.05 & -0.15 \\
$^{151}$Sm & 1.4 & 39.0 & -0.30 & 0.3 & 0.15 & -0.18 \\
$^{153}$Sm & 1.9 & 35.0 & -0.18 & 0.7 & 2.5 & -0.50 \\
$^{155}$Sm & 1.6 & 22.5 & -0.28 & 0.58 & 1.25 & -0.30 \\
$^{195}$Pt & 0.5 & 0.6 & -0.36 & 0.6 & 0.5 & -0.85 \\
$^{195}$Au & 0.6 & 1.45 & -0.35 & 0.65 & 2.0 & -0.33 \\
\hline\hline
\end{tabular*}
\end{center}
\end{table}

Once all the parameters of the different building blocks of the IBFM Hamiltonian $\hat H$ 
are fixed by the procedure described above, the Hamiltonian is 
diagonalized in the
spherical basis $|j, L, \alpha, J\rangle$ \cite{PBOS}, where
$\alpha=(n_d,\nu,n_{\Delta})$ represents a 
generic notation for the quantum numbers of the U(5) symmetry in the IBM 
\cite{IBM}, $L$ and $J$ are the angular momentum of the boson and
the total angular momentum of the coupled boson-fermion system,
respectively. They satisfy the standard triangular selection rule $|L-j|\leq J\leq L+j$. The wave 
functions resulting from the diagonalization of $\hat H$ are  used to 
compute the $B(E2)$ and $B(M1)$ transition rates as well as
spectroscopic quadrupole and magnetic moments. The electric E2 transition 
operator is taken as the sum of the boson and fermion parts
\begin{equation} \label{TE2Rayner}
\hat T^{(E2)}=\hat T^{(E2)}_B+\hat T^{(E2)}_F
\end{equation}
with the boson operator  given by 
\begin{equation} \label{TE2RaynerBoson}
T^{(E2)}_B=e_B\hat Q_B
\end{equation}
where $e_B$ is the boson
effective charge and $\hat Q_B$ represents the quadrupole operator defined in
Eq.~(\ref{eq:ibm}) with the same value of the parameter $\chi$.  On the other 
hand, the fermion E2 operator takes the form
\begin{equation}
 \hat T^{(E2)}_F=
-e_F\sum_{jj^{\prime}}
\frac{1}{\sqrt{5}}
\gamma_{jj^{\prime}}
[a^{\dagger}\times\tilde a_{j^{\prime}}]^{(2)},
\end{equation}
with $e_F$ being the fermion effective charge. As in previous studies 
\cite{nomura2016odd,nomura2016qpt},  $e_B$ is
fitted to reproduce the experimental $B(E2; 2^+_1\rightarrow 0^+_1)$
value of the corresponding even-even boson-core nuclei while  $e_F$ is 
taken as $e_F=e_B$ for all the
considered odd-mass nuclei.

In the same fashion, the magnetic M1 transition operator is given by
\begin{equation}
 \hat T^{(M1)}=\sqrt{\frac{3}{4\pi}}(\hat T^{(M1)}_B + \hat T^{(M1)}_F)
\end{equation}
where the M1 boson operator is proportional to the boson angular momentum
operator  $\hat T^{(M1)}_B=g_B\hat L$ with the gyro-magnetic factor $g_B=\mu_{2^+_1}/2$
given in terms of the magnetic moment $\mu_{2^+_1}$ of the $2^+_1$ state of the
even-even nucleus.  The corresponding experimental value is used for this
quantity. 
The  fermion part is written as
$\hat T^{(M1)}_F$ takes the form \cite{scholten1985}
\begin{equation}
 \hat T^{(M1)}_F=-\sum_{jj^{\prime}} g_{jj^{\prime}}
 \sqrt{\frac{j(j+1)(2j+1)}{3}}[a^{\dagger}_j\times\tilde a_{j^{\prime}}]^{(1)},
\end{equation}
with the coefficients $ g_{jj^{\prime}}$ given by
\begin{equation}
 g_{jj^{\prime}}
=\left\{\begin{array}{ll}
\frac{(2j-1)g_l + g_s}{2j} & (j=j^{\prime}=l+\frac{1}{2}) \\
  \frac{(2j+3)g_l - g_s}{2(j+1)} & (j=j^{\prime}=l-\frac{1}{2}) \\
	 (g_l-g_s)\sqrt{\frac{2l(l+1)}{j(j+1)(2j+1)(2l+1)}} &
	  (j^{\prime}=j-1; l=l^{\prime}) \\
	\end{array} \right.
\end{equation}
In this expression $l$ represents the orbital angular momentum of the 
single-particle state. Throughout this work, the fermion  $g_l$ and 
$g_s$ gyro-magnetic factors take the usual Schmidt values  $g_l=1.0$ 
$\mu^2_N$ and $g_s=5.58$ $\mu^2_N$ for proton and $g_l=0$ and
$g_s=-3.82$ $\mu^2_N$ for neutron. The $g_s$ is quenched by 30 \% for
both proton and neutron, as in Refs.~\cite{scholten1982,nomura2016odd}.

\subsection{Comparison of the parameters with the ones of the relativistic EDF}


\begin{figure}[htb!]
\begin{center}
\includegraphics[width=0.8\columnwidth]{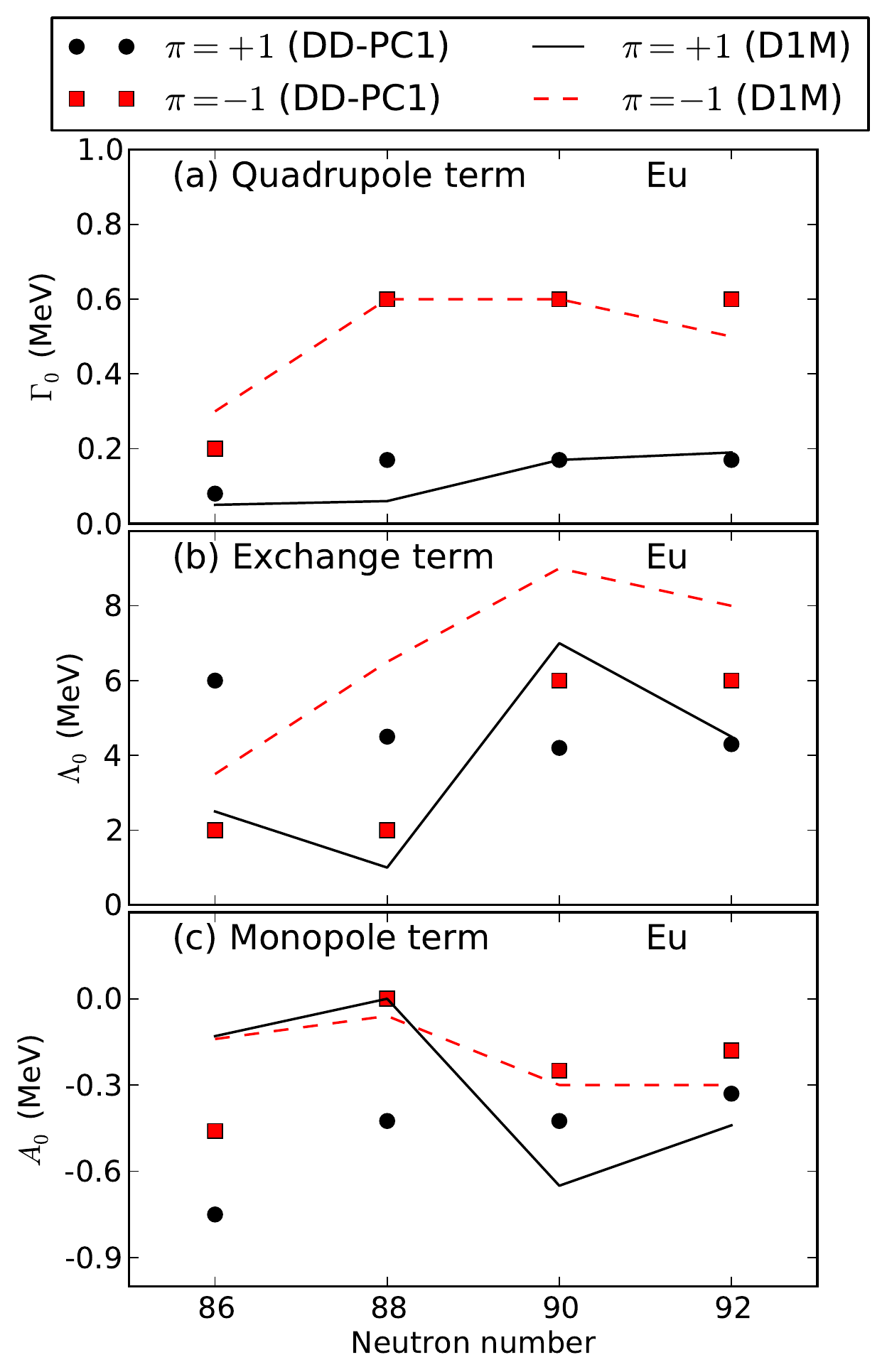}
\caption{(Color online) The coupling constants of the
 boson-fermion interaction term $\hat H_{BF}$ for the odd-A Eu isotopes,
 used in the present study  (denoted by ``D1M'') and in the previous
 calculation in Refs.~\cite{nomura2016odd,nomura2016qpt}. They are shown
 separately for positive- ($\pi=+1$) and negative-parity ($\pi=-1$) states.}
\label{fig:eu-vbf}
\end{center}
\end{figure}


\begin{figure}[htb!]
\begin{center}
\includegraphics[width=0.8\columnwidth]{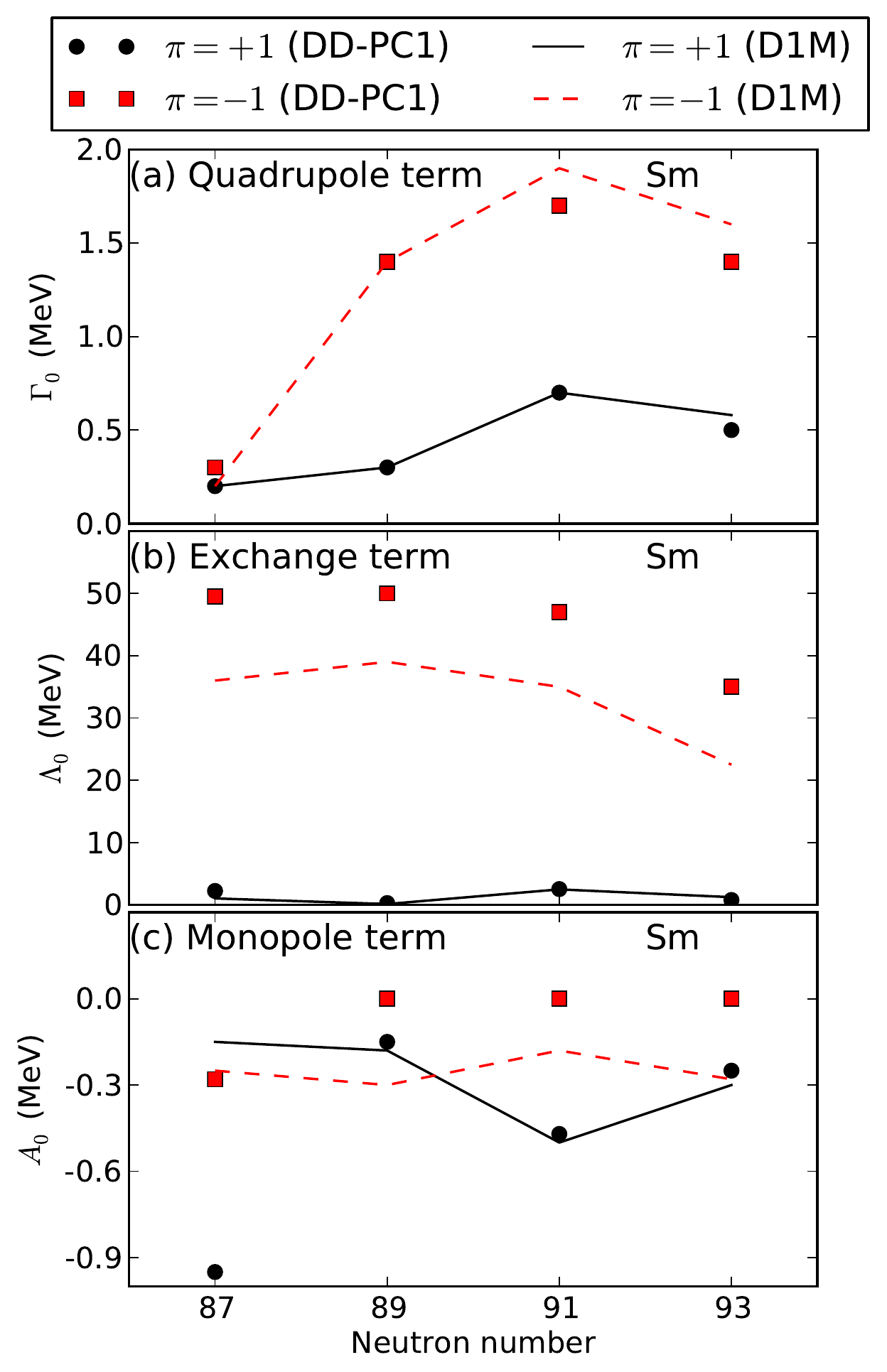}
\caption{(Color online) The same as in
 Fig.~\ref{fig:eu-vbf}, but for the odd-A Sm isotopes.}
\label{fig:sm-vbf}
\end{center}
\end{figure}


\begin{figure}[htb!]
\begin{center}
\includegraphics[width=\columnwidth]{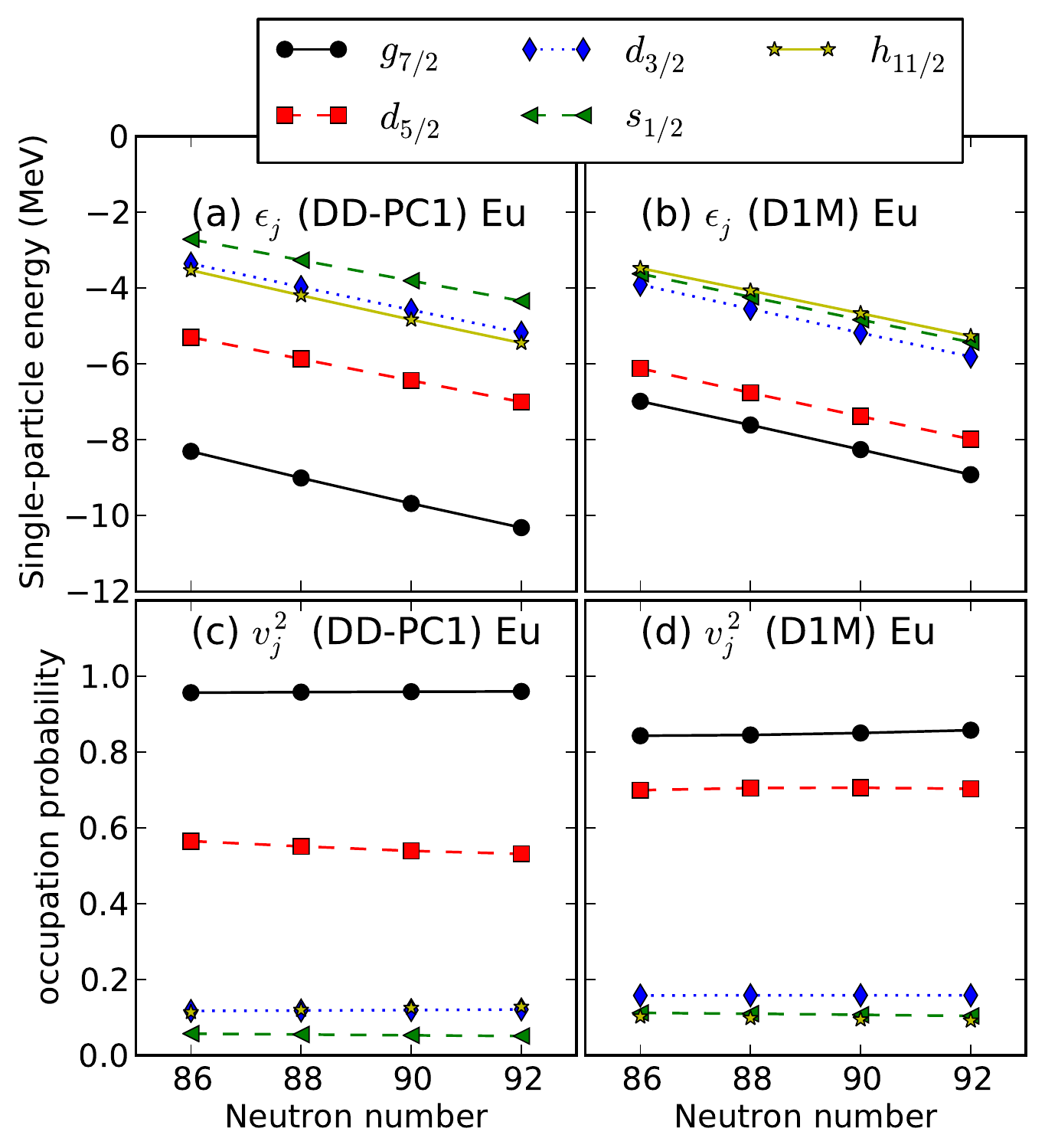}
\caption{(Color online) The spherical single-particle energies
 $\epsilon_j$ and occupation probabilities $v_j^2$ for the odd-A Eu
 isotopes, resulting from the Gogny-D1M HFB calculations, are compared
 with those obtained with the relativistic DD-PC1 EDF in Ref.~\cite{nomura2016qpt}.}
\label{fig:eu-spe}
\end{center}
\end{figure}


\begin{figure}[htb!]
\begin{center}
\includegraphics[width=\columnwidth]{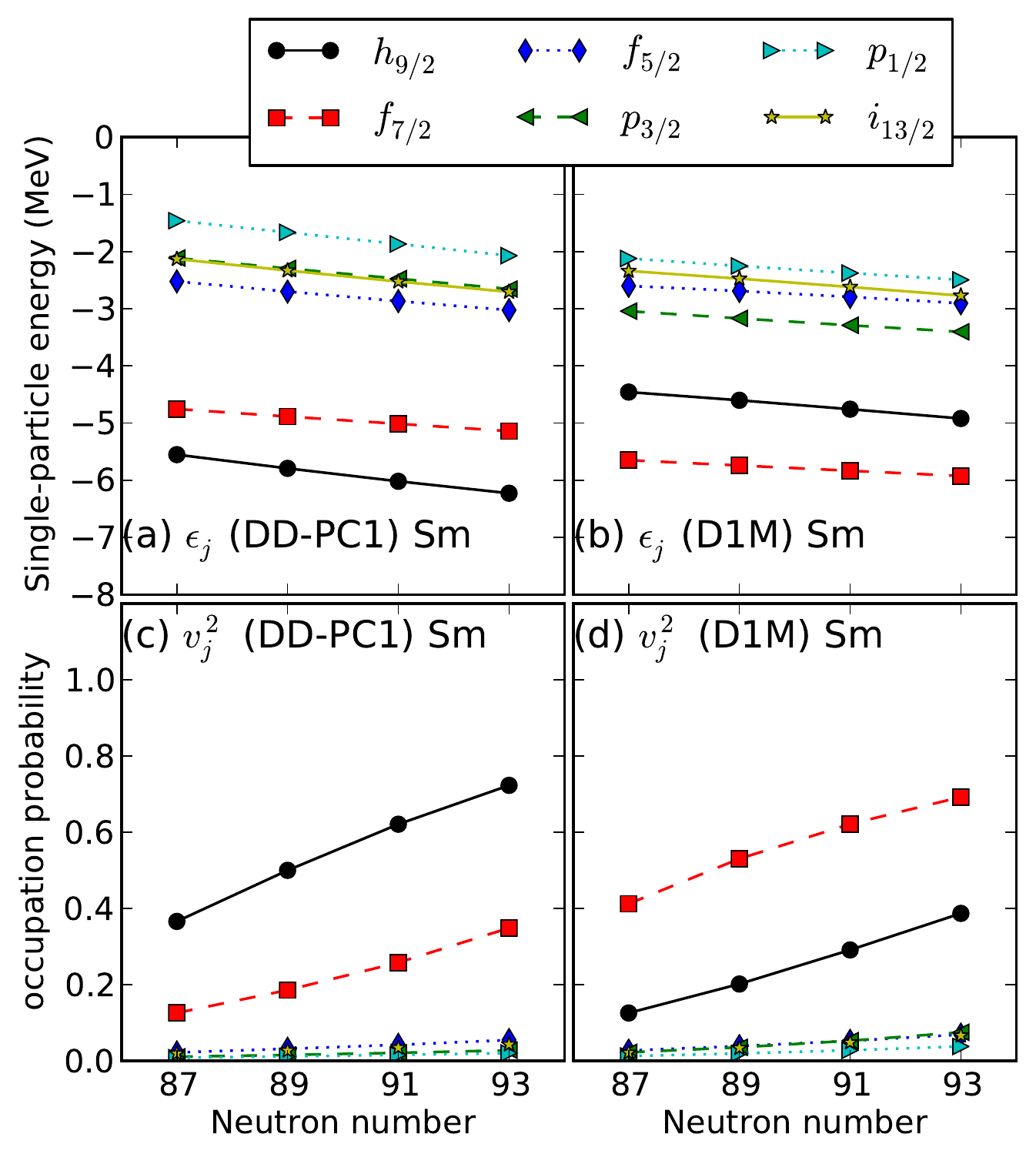}
\caption{(Color online) The same as 
 Fig.~\ref{fig:eu-spe}, but for the odd-A Sm isotopes.}
\label{fig:sm-spe}
\end{center}
\end{figure}

At this point, it is worth to point out some differences found between
the values of the parameters entering the IBFM Hamiltonian, especially
$\hat H_F$ and $\hat H_{BF}$, obtained for
the odd-A Eu and Sm isotopes in the present work as compared to the ones
used in the previous studies of the same isotopes within the
relativistic EDF framework \cite{nomura2016odd,nomura2016qpt}.

Firstly, in Figs.~\ref{fig:eu-vbf} and \ref{fig:sm-vbf}, we plot the strength
parameters $\Gamma_0^\pm$, $\Lambda_0^\pm$ and 
$A_0^\pm$ obtained for the odd-A Eu and Sm isotopes, respectively, from
the Gogny-D1M and relativistic DD-PC1 EDFs.  
As one can observe in Figs.~\ref{fig:eu-vbf}(a) and (b), significant 
discrepancies between the present and previous 
\cite{nomura2016odd,nomura2016qpt} studies are found in the fitted
$\Lambda_0^{\pm}$ and $A_0^{\pm}$ values, especially for lighter
isotopes $^{149,151}$Eu, both quantitatively and qualitatively. In
addition, in Ref.~\cite{nomura2016odd} the monopole term was introduced
only for the $2d_{5/2}$ orbit for the positive-parity states in the
odd-mass Eu isotopes. In this study, on the other hand, we have
introduced the monopole term for all the positive-parity orbitals in a
given isotope, with a common value $A_0^+$. 
On the other hand, the values of the strength parameters
($\Gamma_0^\pm$, $\Lambda_0^\pm$ and $A_0^\pm$) obtained here for 
the odd-A Sm isotopes (see Fig.~\ref{fig:sm-vbf}) are quite
similar to those employed in Ref.~\cite{nomura2016qpt}, except perhaps
for the $A^+_0$ values for $^{149}$Sm (see, Fig.~\ref{fig:sm-vbf}(c))

The observed differences in the boson-fermion strength parameters for
the odd-A Eu between the present and previous \cite{nomura2016odd} studies
could partly originate from the quantitative differences, especially in the
single-particle energies, between the Gogny and relativistic EDFs
quantities entering the fit. 
To illustrate this possibility, we plot in Fig.~\ref{fig:eu-spe} the
spherical single-particle energies and occupation probabilities for the
odd-A Eu isotopes obtained in Ref.~\cite{nomura2016odd} and in the present work. 
As seen in panels (a) and (b) of Fig.~\ref{fig:eu-spe}, the energy gap
between the $1g_{7/2}$ and $2d_{5/2}$ orbitals for the odd-mass Eu
isotopes is, in general, more than 3 MeV in Ref.~\cite{nomura2016odd}
while it is  less than 1 MeV in the Gogny-D1M calculations (see, also
Table~\ref{tab:spe}). 
The values of $\epsilon_j$ and $v_j^2$ obtained for the odd-A Sm
isotopes from the Gogny-D1M and
the relativistic EDFs are plotted in Fig.~\ref{fig:sm-spe}. 

In addition, we observe, in Fig.~\ref{fig:sm-vbf}(b) as well as in
Table~\ref{tab:paraBF}, that the values of $\Lambda^+_0$ for
$^{149-155}$Sm are rather large ($\approx 20-40$ MeV), about a  
factor of ten larger than the $\Lambda_0^{\pm}$ parameters obtained for 
other odd-mass nuclei. They are also larger than the ones employed in 
earlier phenomenological IBFM calculations for other isotopic chains 
\cite{IBFM-Book}. The reason for the large $\Lambda_0^+$ values in the 
odd-mass Sm nuclei is the nearly vanishing 
$\beta_{jj^{\prime\prime}}\beta_{j^{\prime}j^{\prime\prime}}$ factors in the 
strength parameters $\Lambda^{j^{\prime\prime}}_{jj^{\prime}}$ (see 
Eq.~(\ref{eq:exchange})) consequence of the too small $v^2_{i_{13/2}}$ 
values used (see Table~\ref{tab:vv} and Fig.~\ref{fig:sm-spe}(d)). 
We note, that even larger values of $\Lambda^+_0$ for the odd-A Sm
isotopes than the present ones have been obtained in 
the case where relativistic DD-PC1 EDF was used \cite{nomura2016qpt}
(see, Fig.~\ref{fig:sm-vbf}(b)). 
In Ref.~\cite{nomura2016qpt}, too small $v_{i_{13/2}}^2$ values for the
odd-A Sm isotopes were also obtained, similarly to the present work (see,
Fig.~\ref{fig:sm-spe}(d)).


\section{Results for the even-even core nuclei\label{sec:pes}}



\begin{figure*}[htb!]
\begin{center}
\includegraphics[width=\linewidth]{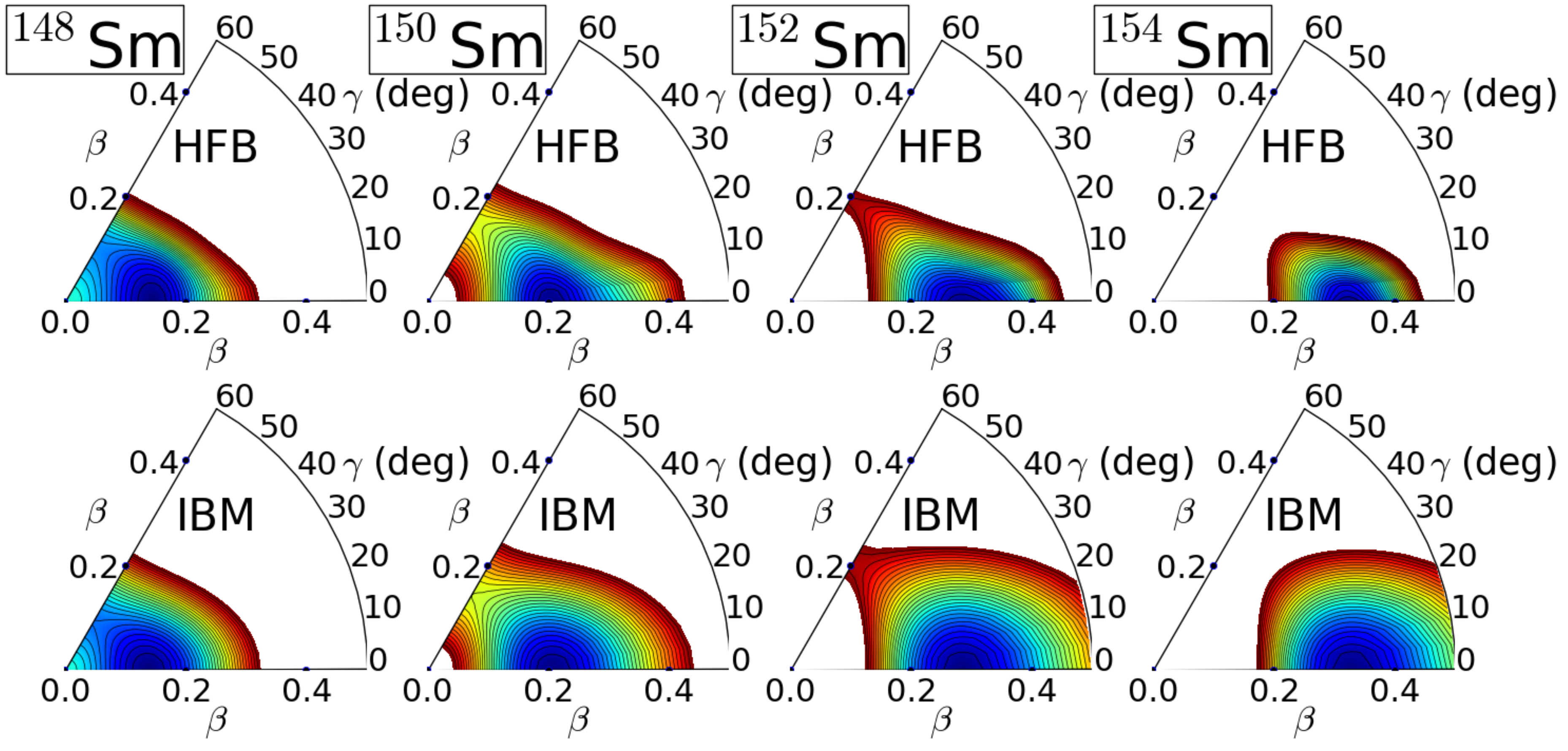}
\caption{(Color online) The Gogny-D1M (upper row) and mapped IBM (lower row) energy surfaces 
 for the even-even nuclei $^{148-154}$Sm are plotted up to 3 MeV above the absolute
 minimum. The difference between neighboring contours is 100 keV.}
\label{fig:sm-pes}
\end{center}
\end{figure*}


\begin{figure}[htb!]
\begin{center}
\includegraphics[width=\linewidth]{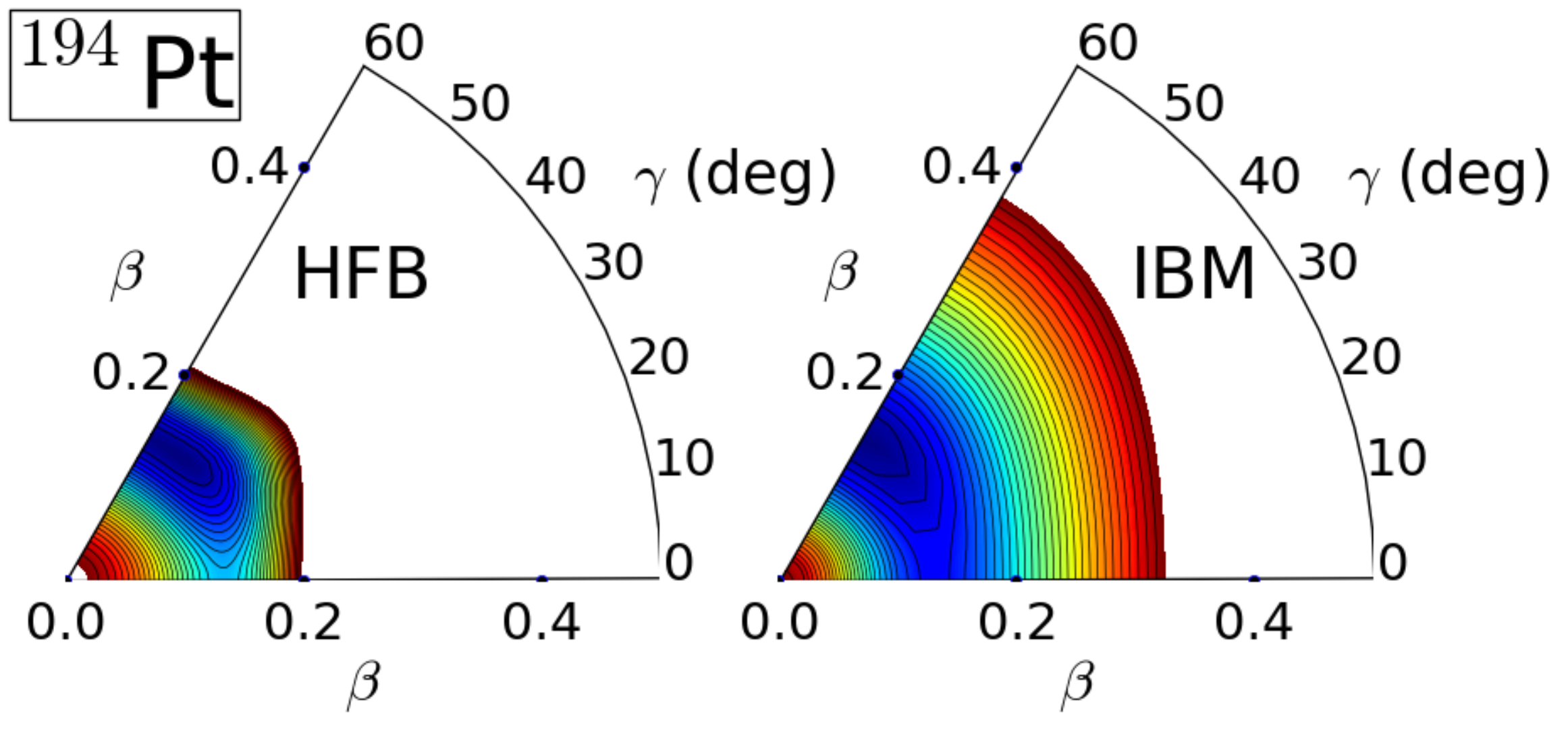}
\caption{(Color online) Same as in  Fig.~\ref{fig:sm-pes}, but for 
 $^{194}$Pt.}
\label{fig:pes-pt}
\end{center}
\end{figure}


\begin{figure}[htb!]
\begin{center}
\includegraphics[width=\linewidth]{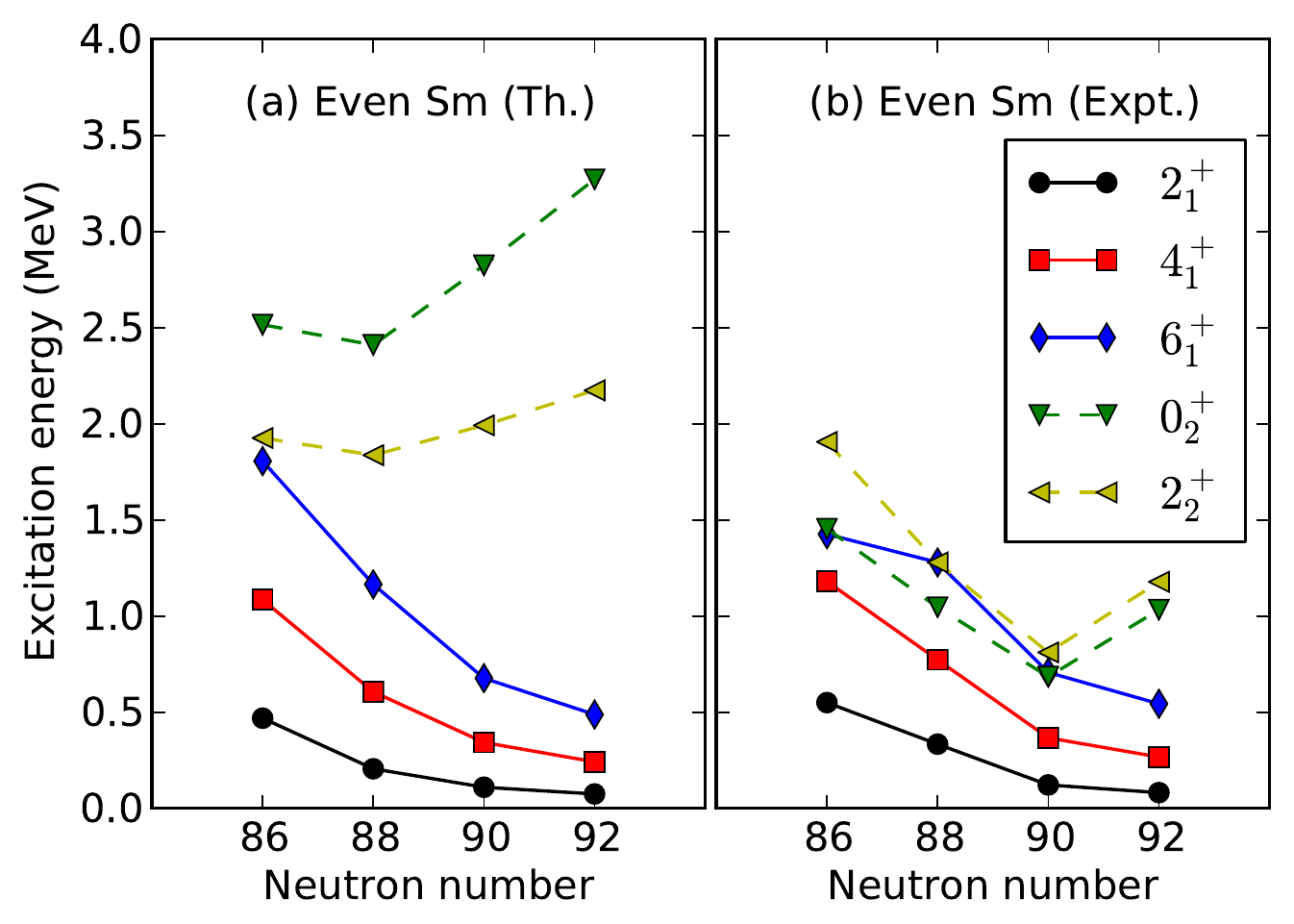}
\caption{(Color online) The  low-lying states in $^{148-154}$Sm are plotted 
as functions of the neutron number $N$. Experimental data have been taken 
from Ref.~\cite{data}.}
\label{fig:evensm-level}
\end{center}
\end{figure}


\begin{figure}[htb!]
\begin{center}
\includegraphics[width=\linewidth]{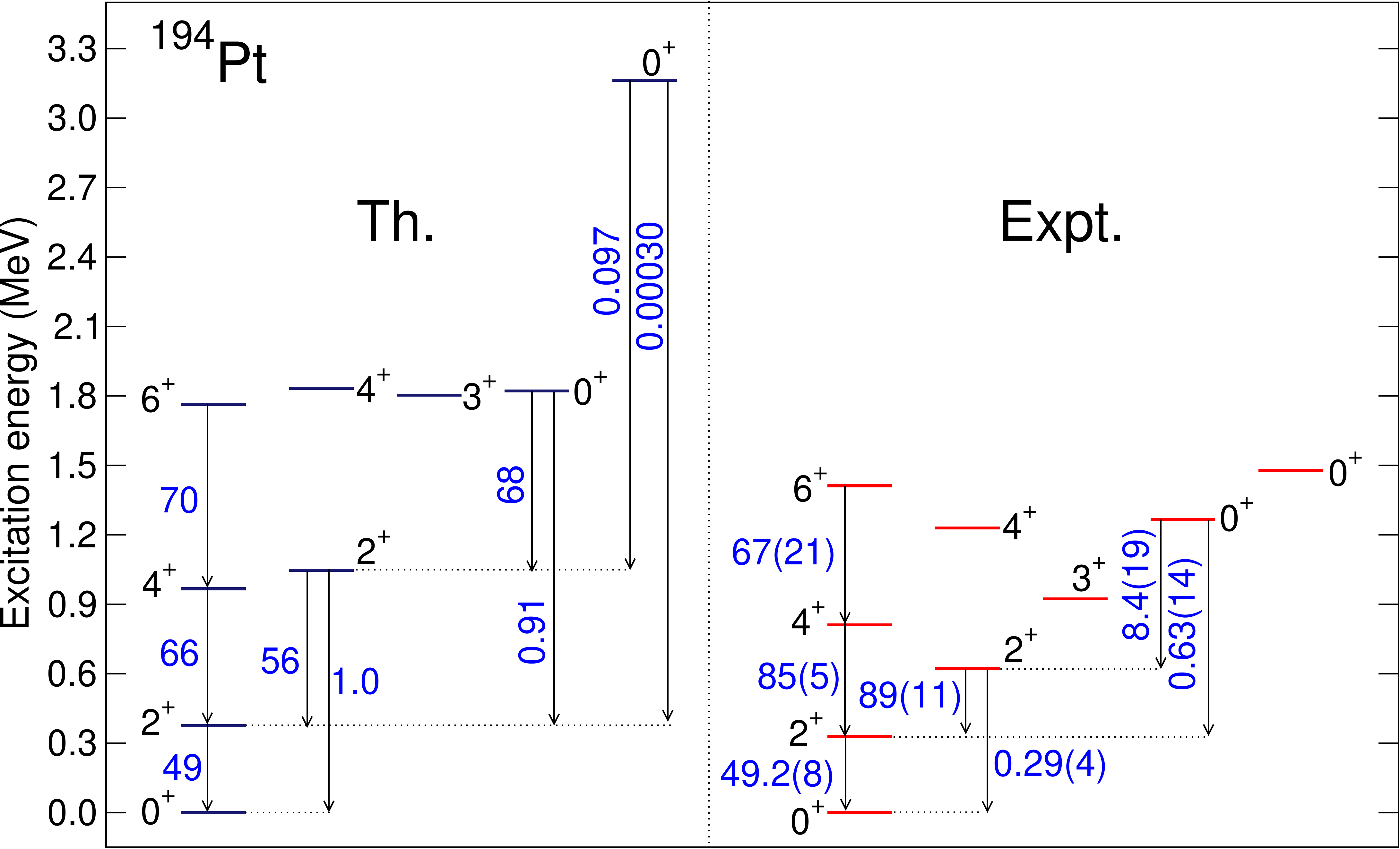}
\caption{(Color online) Energy spectra of $^{194}$Pt. The numbers along 
arrows are $B(E2)$ transition strengths in Weisskopf units. 
Experimental data have been taken from Ref.~\cite{data}.}
\label{fig:194pt}
\end{center}
\end{figure}

The Gogny-D1M and mapped IBM energy surfaces obtained for the even-even 
boson-core nuclei $^{148-154}$Sm are plotted in Fig.~\ref{fig:sm-pes}. 
Those surfaces illustrate the transition between nearly spherical and 
axially-deformed shapes \cite{cejnar2010}. In the case of $^{148}$Sm 
the HFB surface exhibits a weakly deformed minimum around $\beta=0.15$. 
The nucleus $^{150}$Sm displays a sharper potential in both $\beta$ and 
$\gamma$ directions with a minimum around $\beta=0.2$. The minimum of 
the HFB surface obtained for $^{152}$Sm is even sharper, especially 
along the $\gamma$ direction but it looks softer in $\beta$ than in  
$^{150}$Sm. This softening of the potential in $\beta$ agrees well with 
a key feature of a transitional nucleus associated with the X(5) 
critical-point symmetry \cite{X5,casten2001}. Finally, the nucleus  
$^{154}$Sm exhibits the most pronounced prolate deformation with 
$\beta\approx 0.35$. The mapped IBM surfaces, plotted in the lower row 
of the figure, reproduce nicely the Gogny-HFB ones for each nucleus, 
exception made of the fact that, due to the limited IBM configuration 
space used in this work, far away from the minimum the IBM surfaces tend to 
be flatter than the Gogny-HFB ones. In Fig.~\ref{fig:pes-pt}, we have plotted 
the Gogny-D1M and IBM energy contour plots for $^{194}$Pt. Both 
surfaces exhibit a typical $\gamma$ softness with a weakly-deformed 
oblate minimum at $\beta\approx 0.15$

The low-energy levels, resulting from the diagonalization of the IBM 
Hamiltonian, are plotted in Fig.~\ref{fig:evensm-level} as functions of 
the neutron number $N$ for the isotopes $^{148-154}$Sm. They are 
compared with the available experimental data \cite{data}. The 
calculations reproduce reasonably well the experimental trends of the 
low-lying energy levels and suggest the transition from a 
vibrational-like spectrum at $N=86$ to the typical rotational-like 
spectrum at $N=92$. The overestimation of the energies of the $0^+_2$ 
and $2^+_2$ states has also been found in earlier calculations within 
the fermion-to-boson mapping   procedure \cite{nomura2008,nomura2010}. 
A reason for that could be the restricted model space of the IBM and/or to the 
fact that the shape of the Gogny-HFB energy surfaces 
around the minimum have too large curvatures  in both the $\beta$ and $\gamma$ directions which 
require a large value of the quadrupole-quadrupole interaction strength 
$\kappa$ (see Eq.~(\ref{eq:ibm})) in the IBM Hamiltonian. The large $\kappa${}
values push up the non-yrast energy levels. 

The low-energy spectrum obtained for  $^{194}$Pt  is shown in 
Fig.~\ref{fig:194pt}. It exhibits several features of 
$\gamma$-softness or the O(6) symmetry \cite{IBM}, i.e., the energy 
ratio $R_{4/2}=E(4^+_1)/E(2^+_1)=2.57$, the multiplets ($4^+_1,2^+_2$) 
and ($6^+_1, 4^+_2, 3^+_1$), the large $B(E2; 2^+_2\rightarrow 2^+_1)$ 
transition strength of the same order of magnitude as the $B(E2; 
4^+_1\rightarrow 2^+_1)$ one and the selection rule of the E2 decay 
from the $0^+_2$ to $2^+_{1,2}$ state. When compared with the experimental
data, the spectrum looks rather stretched.

\section{Spectroscopic properties of odd-$A$ Eu and Sm nuclei\label{sec:eusm}}



\begin{figure}[htb!]
\begin{center}
\includegraphics[width=\linewidth]{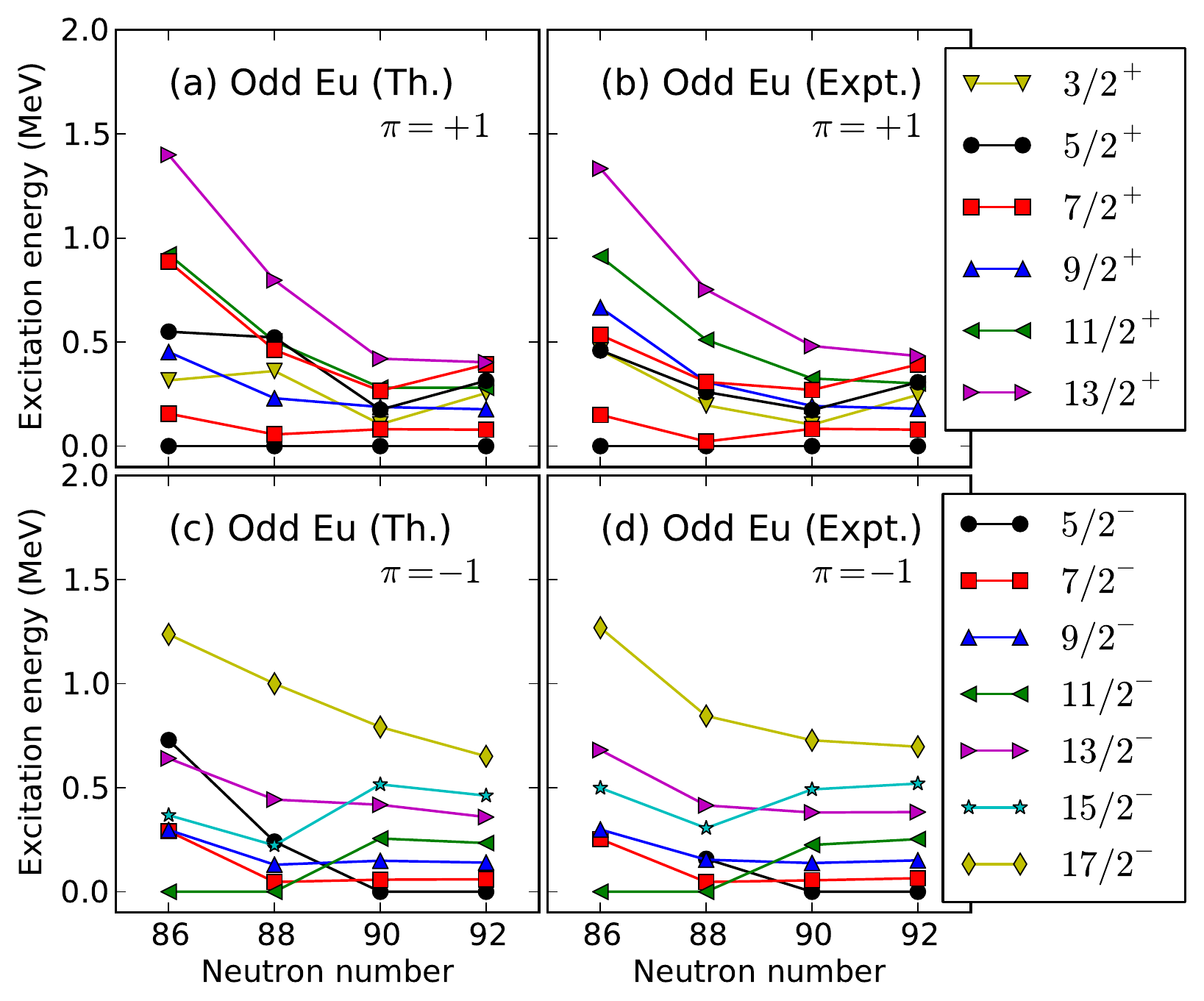}
\caption{(Color online) The low-lying positive- ($\pi=+1$) and
 negative-parity ($\pi=-1$) states in the odd-mass isotopes $^{149-155}$Eu 
are plotted as functions of the neutron number $N$. Experimental 
data have been taken from Ref.~\cite{data}.} 
\label{fig:eu-level}
\end{center}
\end{figure}


\begin{figure}[htb!]
\begin{center}
\includegraphics[width=\linewidth]{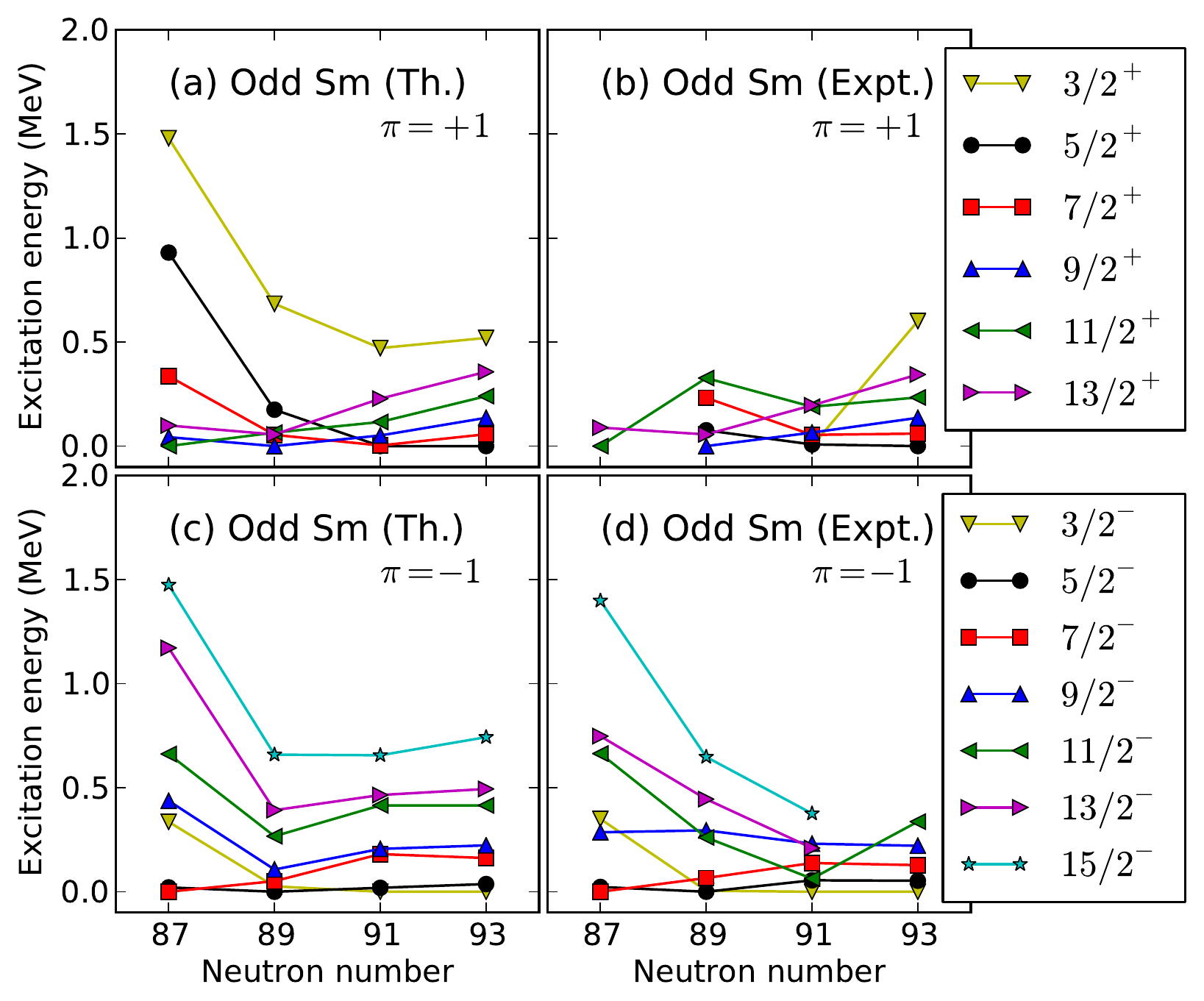}
\caption{(Color online) Same as in Fig.~\ref{fig:eu-level}, but for the
odd-mass Sm isotopes.}
\label{fig:sm-level}
\end{center}
\end{figure}


\begin{figure}[htb!]
\begin{center}
\includegraphics[width=0.7\linewidth]{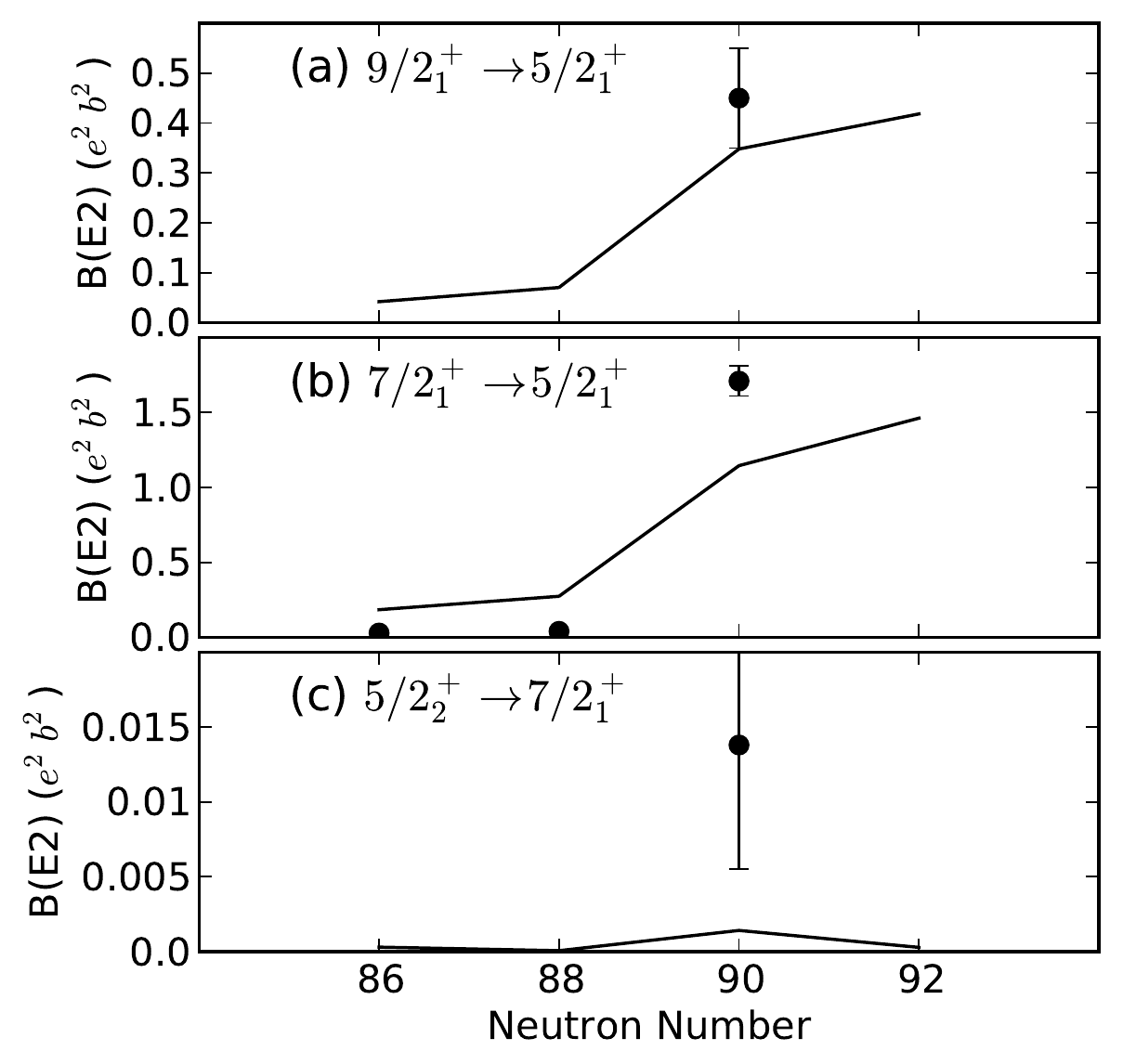}
\caption{The $B(E2; {9/2}^+_1\rightarrow
{5/2}^+_1)$, $B(E2; {7/2}^+_1\rightarrow {5/2}^+_1)$ and $B(E2;
{5/2}^+_2\rightarrow {7/2}^+_1)$ transition strengths obtained for the  odd-mass
isotopes  $^{149-155}$Eu are depicted as  functions of the neutron number $N$.  
Experimental data have been taken from Ref.~\cite{data}.}
\label{fig:eu-e2}
\end{center}
\end{figure}


\begin{figure}[htb!]
\begin{center}
\includegraphics[width=0.7\linewidth]{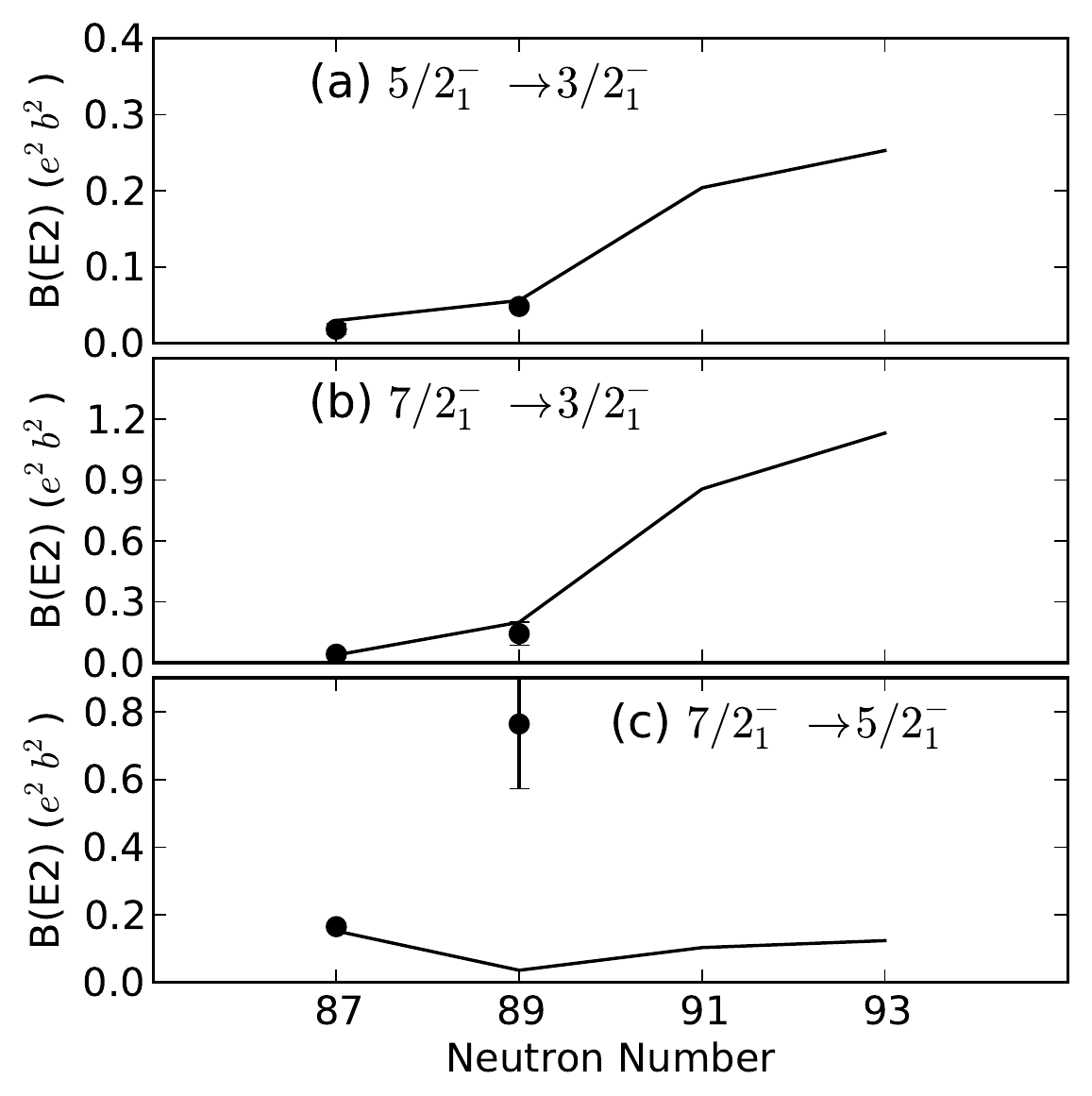}
\caption{The $B(E2; {5/2}^-_1\rightarrow
{3/2}^-_1)$, $B(E2; {7/2}^-_1\rightarrow {3/2}^-_1)$ and $B(E2;
{7/2}^-_1\rightarrow {5/2}^-_1)$ transition strengths 
obtained for the odd-mass isotopes  $^{149-155}$Sm are depicted  
as functions of the neutron number $N$. Experimental data have been 
taken from Ref.~\cite{data}.}
\label{fig:sm-e2}
\end{center}
\end{figure}


\begin{figure}[htb!]
\begin{center}
\includegraphics[width=\linewidth]{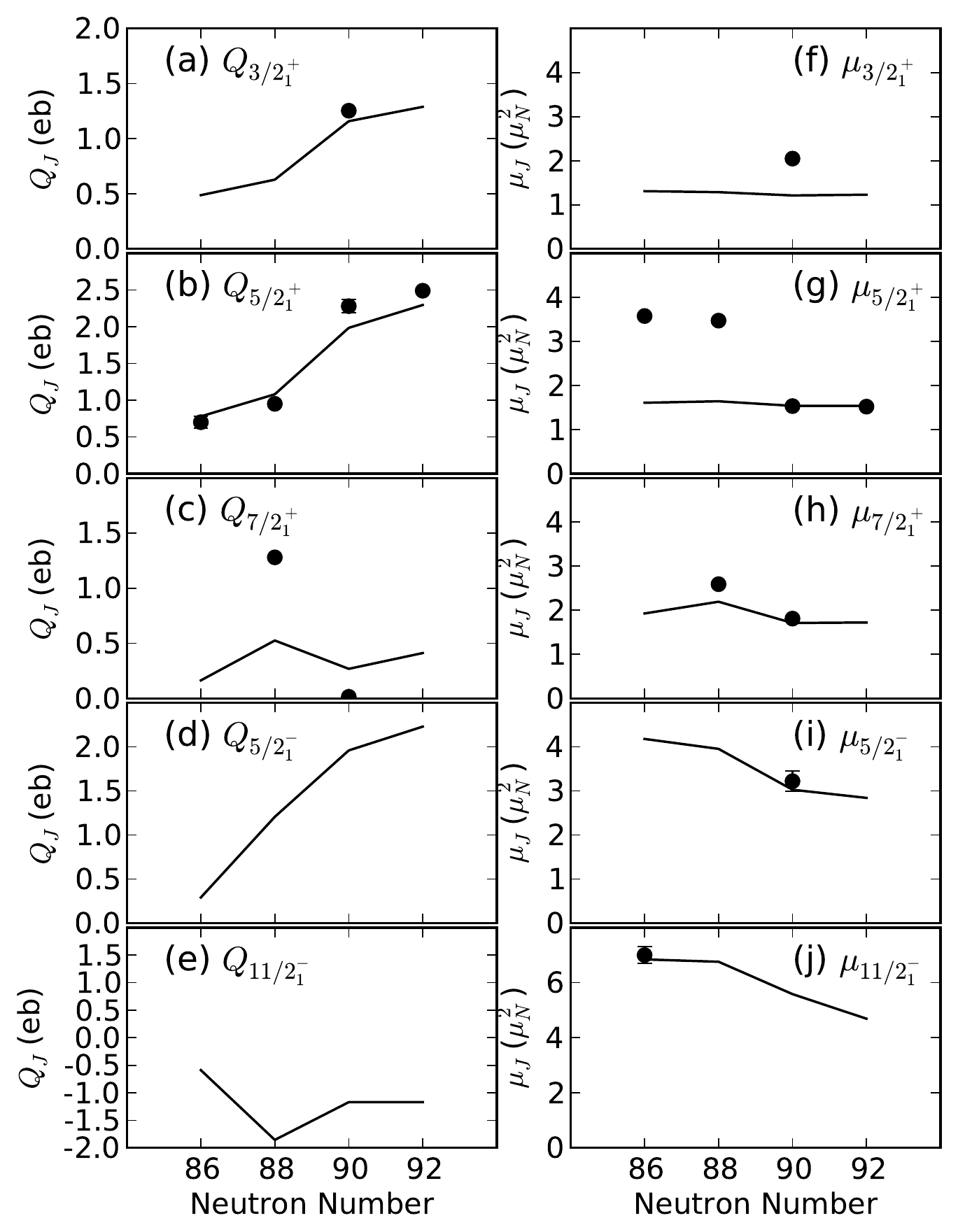}
\caption{The spectroscopic quadrupole $Q_J$ (in $e$b units) and magnetic
$\mu_J$ (in $\mu_N^2$ units) moments 
obtained for the odd-mass isotopes $^{149-155}$Eu
are plotted  as functions of the neutron number  $N$. 
Experimental data, represented by dots, have been taken from 
Ref.~\cite{stone2005}.}
\label{fig:eu-mom}
\end{center}
\end{figure}


\begin{figure}[htb!]
\begin{center}
\includegraphics[width=\linewidth]{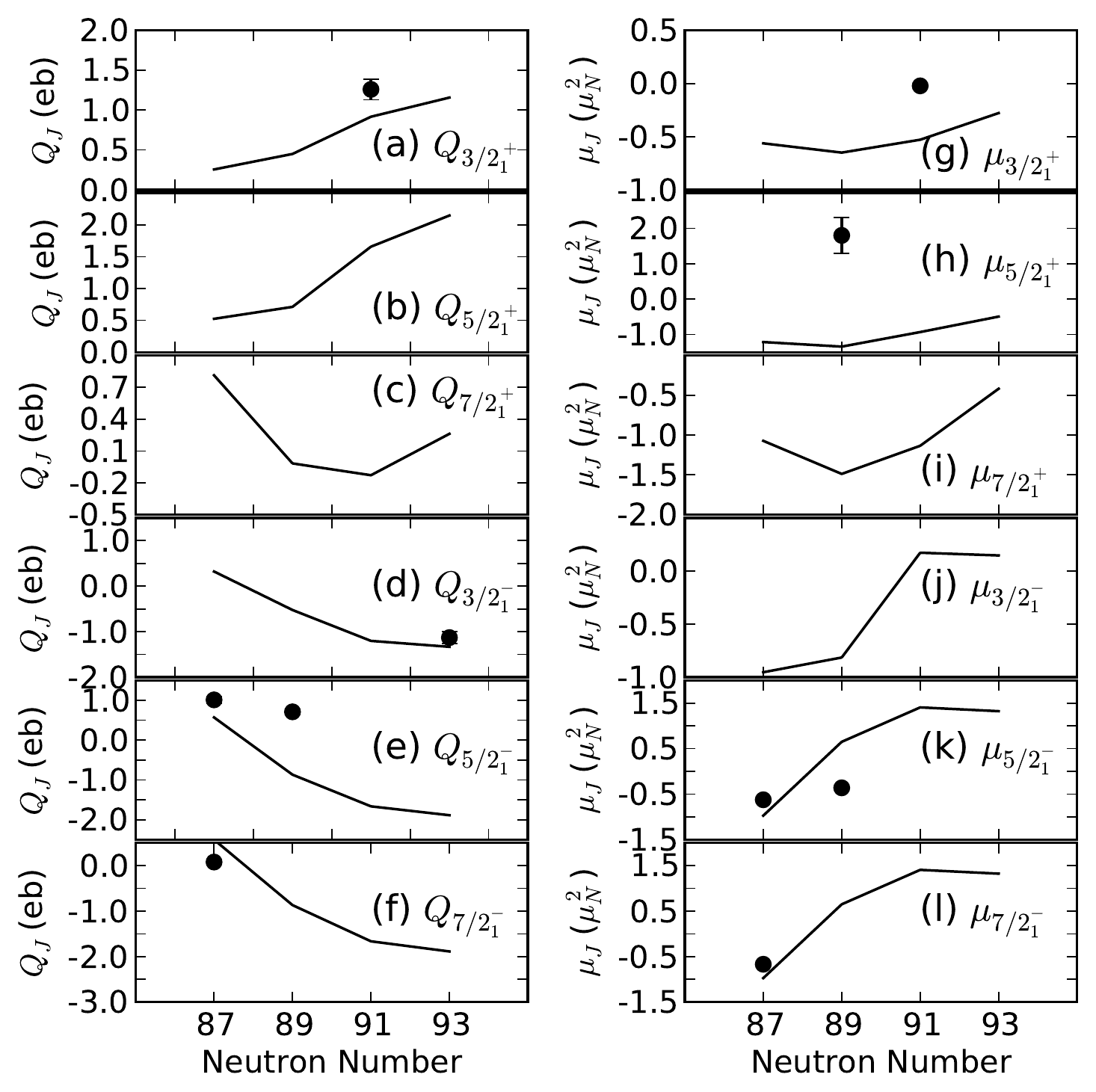}
\caption{Same as in  Fig.~\ref{fig:eu-mom}, but for the odd-mass isotopes
$^{149-155}$Sm.}
\label{fig:sm-mom}
\end{center}
\end{figure}

Having demonstrated that the mapped IBM Hamiltonian gives a reasonable 
description of the even-even (boson-core) nuclei, we now turn our 
attention to the considered  odd-mass nuclei. In 
Fig.~\ref{fig:eu-level} we have depicted the  low-lying positive- and 
negative-parity states obtained for the studied odd-$A$ Eu isotopes as 
functions of neutron number $N$. They are compared with the available 
experimental data \cite{data}. A reasonable good agreement between the 
theoretical predictions and the experimental values is observed. The 
compression of the  positive- and negative-parity  levels, as functions 
of the neutron number, correlates well with the lowering of the yrast 
levels in  the even-even nuclei (see, Fig.~\ref{fig:evensm-level}). It 
can be  regarded as  a signature of the structural evolution from the 
nearly spherical to the well-deformed regime. Another signature of 
shape transition, specific to the odd-mass nuclei, is the change in the 
angular momentum of the ground state in the case of 
negative-parity configurations from $N=88$ to 90 [see, panels (c) and 
(d)]. Our results suggest that for $^{149,151}$Eu the ${11/2}^-_1$ is 
the ground state for negative parity which is weakly coupled to the boson 
core (represented by $^{148,150}$Sm) that exhibits a moderate 
deformation (see, Fig.~\ref{fig:sm-pes}). At $N=90$ and 92, the 
coupling between the odd proton and boson-core nucleus becomes stronger 
and the regular rotational band built on the ${5/2}^-_1$ state, that 
follows the $\Delta J=1$ systematics of the strong coupling limit, 
emerges. In the case of the positive-parity states [panels (a) and 
(b)], the ${5/2}^+$ one remains the ground state. Nevertheless, both 
theoretically and experimentally, the ${5/2}^+_1$ and ${7/2}^+_1$ states 
are rather close in energy  at $N=88$. This feature can be regarded as 
a signature of the structural change taking place around $N=88$. 

The theoretical and experimental spectra for the odd-mass Sm isotopes 
are plotted in Fig.~\ref{fig:sm-level}. As can be seen, our 
calculations provide a reasonable description of the experimental data. 
A change in the  spin of the ground state  is observed at $N=91$. On 
the other hand, the levels are more compressed from $N=87$ towards 89 
or 91 [panels (a), (c), and (d)] suggesting a structural change in 
those odd-mass systems. The experimental data  indicate that for the 
transitional isotope $^{153}$Sm many levels are found below 
$\approx$300 keV. They also reveal  [panels (b) and (d)] a more regular 
rotational-like band that exhibits the $\Delta J=1$ systematics of the 
strong coupling regime. Our results suggest that for both parities the 
rotational-like band appears already at $N=91$ [panels (a) and (c)]. 
However, the predicted negative-parity levels for the $N=91$ and 93 
nuclei [see panel (c)], look rather irregular with the staggering pattern 
$({3/2}^-_1,{5/2}^-_1),({7/2}^-_1,{9/2}^-_1),({11/2}^-_1,{13/2}^-_1),\ldots$. 
We notice here that the Gogny-D1M energy systematic obtained for the 
odd-mass Eu and Sm nuclei is similar to the results found within the 
relativistic EDF framework \cite{nomura2016odd,nomura2016qpt}.
 
In addition to the energy spectra, the electromagnetic transition rates 
also provide signatures of the structural evolution in the considered 
odd-mass nuclei. The $B(E2; {9/2}^+_1\rightarrow {5/2}^+_1)$, $B(E2; 
{7/2}^+_1\rightarrow {5/2}^+_1)$ and $B(E2; {5/2}^+_2\rightarrow 
{7/2}^+_1)$  transition probabilities obtained for the odd-mass Eu 
isotopes are shown in Fig.~\ref{fig:eu-e2} as functions of the neutron 
number. There is a sharp increase in the $B(E2; {9/2}^+_1\rightarrow 
{5/2}^+_1)$ and $B(E2; {7/2}^+_1\rightarrow {5/2}^+_1)$ strengths in going from $N=88$ to 
90 suggesting a sudden structural change from  $^{151}$Eu to 
$^{153}$Eu, especially in the case of the  ${7/2}^+_1\rightarrow 
{5/2}^+_1$ transition. However, our calculations underestimate the experimental  
$B(E2; {5/2}^+_2\rightarrow {7/2}^+_1)$ value although  they still 
exhibit a small peak at $N=90$, where the shape transition occurs.  

The $B(E2; {5/2}^-_1\rightarrow {3/2}^-_1)$, $B(E2; 
{7/2}^-_1\rightarrow {3/2}^-_1)$ and $B(E2; {7/2}^-_1\rightarrow 
{5/2}^-_1)$ transition strengths obtained for the odd-mass isotopes 
$^{149-155}$Sm are plotted in Fig.~\ref{fig:sm-e2} as functions of $N$. 
Our Gogny-D1M  calculations reproduce reasonably well  the available  
data for both the $B(E2; {5/2}^-_1\rightarrow {3/2}^-_1)$ [panel (a)] 
and $B(E2; {7/2}^-_1\rightarrow {3/2}^-_1)$ [panel (b)] strengths in 
$^{149}$Sm and $^{151}$Sm. Experimental data are not available for the 
heavier $N=91$ and/or $N=93$ isotopes.

A sharp rise of the $B(E2; {5/2}^-_1\rightarrow {3/2}^-_1)$ and $B(E2; 
{7/2}^-_1\rightarrow {3/2}^-_1)$ values is predicted from $N=89$ to 91. 
Such a behavior, as in the case of the odd-mass Eu isotopes (see, 
Fig.~\ref{fig:eu-e2}),  points to a spherical-to-deformed  shape 
transition in the corresponding  even-even Sm nuclei. However, the 
predicted $B(E2; {7/2}^-_1\rightarrow {5/2}^-_1)$ transition rate 
[panel (c)] does not exhibit a clear signature of the rapid structural 
change as in the other two cases [panels (a) and (b)]. In particular, 
our calculations underestimate the experimental $B(E2; 
{7/2}^-_1\rightarrow {5/2}^-_1)$ value for $^{151}$Sm. The reason for 
the disagreement is that the computed  wave functions for the 
${5/2}^-_1$ and ${7/2}^-_1$ states are very different in nature, i.e., 
the former is mainly  composed  of $f_{5/2}$ (29 \%) and $h_{9/2}$ (45 
\%) configurations, whereas the latter is mostly made of the $f_{7/2}$ 
configuration (79 \%). 

Another signature of the shape transition already mentioned can be 
found in Fig.~\ref{fig:eu-mom} where the spectroscopic quadrupole 
$Q_J$ and magnetic $\mu_J$ moments  of the ${3/2}^+_1$, ${5/2}^+_1$, 
${7/2}^+_1$, ${5/2}^-_1$ and ${11/2}^-_1$ states   are shown for  
$^{149-155}$Eu. Similar to the $B(E2)$ transition rates, the  $Q_J$ 
values in panels (a) to (e) of the figure exhibit a dramatic change 
around $N=88$ and $N=90$. The agreement between the theoretical and 
experimental \cite{data,stone2005} $Q_J$ values is also very 
reasonable. On the other hand, the  $\mu_J$ values, plotted in panels 
(f) to (j), seem to be less sensitive to $N$ than the  $Q_J$ ones. One 
observes a fair agreement between the calculated and experimental 
$\mu_J$ values, exception made of the substantial disagreement of the 
$\mu_{{5/2}^+_1}$ values at $N=86$ and $N=88$. 
 
In Fig.~\ref{fig:sm-mom}, we have depicted the $Q_J$ and $\mu_J$ 
moments for the ${3/2}^{\pm}_1$, ${5/2}^{\pm}_1$ and ${7/2}^{\pm}_1$ 
states in the case of the odd-mass isotopes $^{149-155}$Sm. As in 
Fig.~\ref{fig:eu-mom}, the predicted $Q_J$ values [panels (a) to (f)] 
exhibit a rapid change around $N=89$ or $N=91$, where the shape transition 
occurs. Notice, that the sign of  $Q_{{3/2}^-_1}$ for $^{155}$Sm is not 
known experimentally \cite{data,stone2005} though it is assumed to be 
negative for consistency with the calculated one. On the other hand, 
many of the predicted $\mu_J$ values [panels (g) to (l)] also exhibit a 
significant change around the transitional system with $N=89$ or $N=91$. 
In many cases, however, the sign of the corresponding $\mu_J$ moments is the opposite 
to the experimental one.


\section{Spectroscopic properties of $^{195}$Pt and $^{195}$Au\label{sec:ptau}}



\begin{figure}[htb!]
\begin{center}
\includegraphics[width=\linewidth]{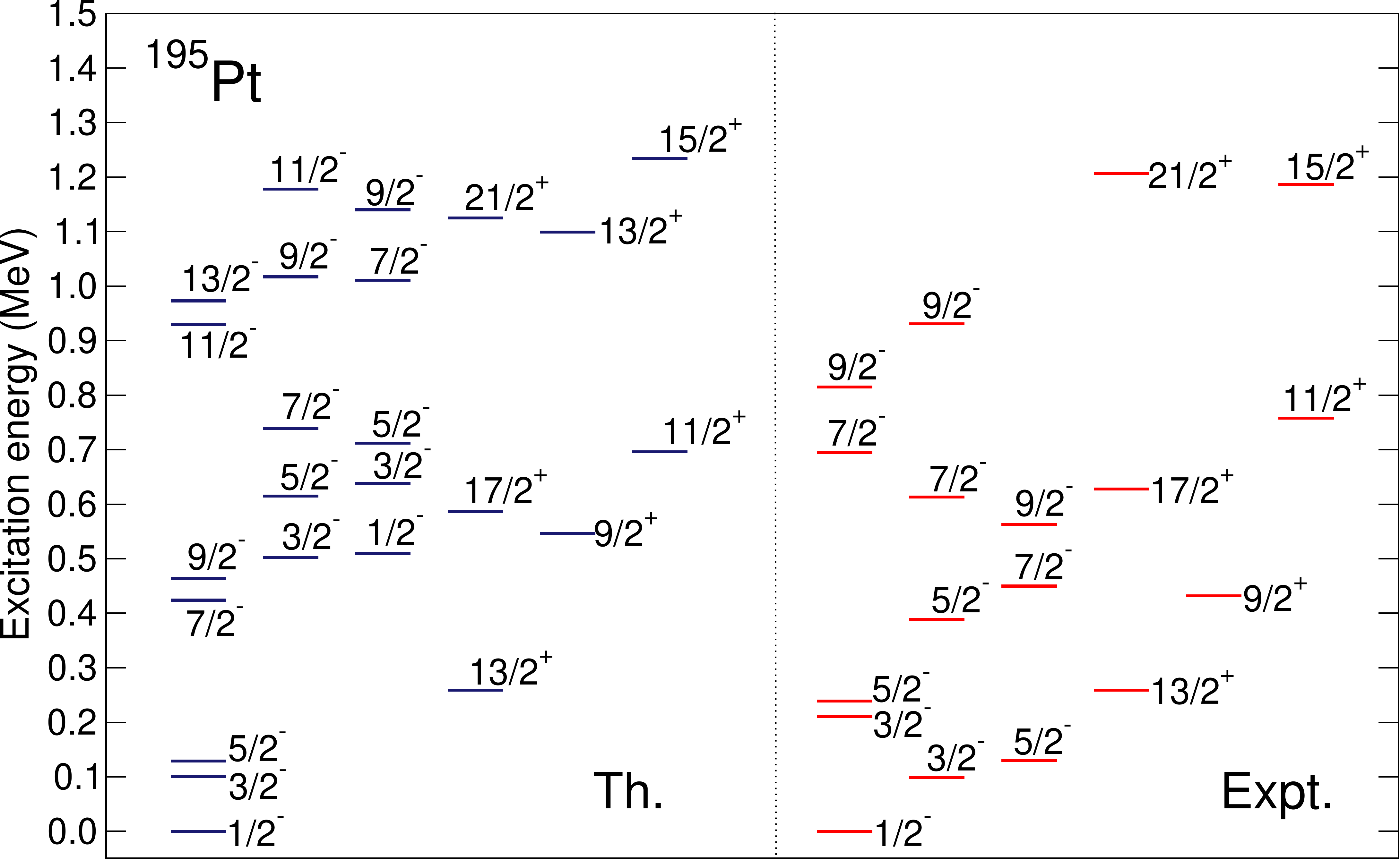}
\caption{(Color online) The three lowest  positive- and negative-parity
 bands in $^{195}$Pt. Experimental data have been taken from Ref.~\cite{data}.}
\label{fig:195pt}
\end{center}
\end{figure}

\begin{table}[hb!]
\caption{\label{tab:195pt-frac} 
Amplitudes (in per cent) of the  negative-parity states of $^{195}$Pt shown in 
Fig.~\ref{fig:195pt}
when expressed in the single particle basis of the $3p_{1/2}$,
$3p_{3/2}$, $2f_{5/2}$, $2f_{7/2}$ and $1h_{9/2}$ orbitals.}
\begin{center}
\begin{tabular*}{\columnwidth}{p{1.33cm}p{1.33cm}p{1.33cm}p{1.33cm}p{1.33cm}p{1.33cm}}
\hline\hline
\textrm{$J^{\pi}$} &
\textrm{$3p_{1/2}$}&
\textrm{$3p_{3/2}$}&
\textrm{$2f_{5/2}$}&
\textrm{$2f_{7/2}$}&
\textrm{$1h_{9/2}$} \\
\hline
${1/2}^{-}_{1}$ & 67 & 12 & 12 & 5 & 4 \\
${1/2}^{-}_{2}$ & 0 & 46 & 39 & 8 & 7 \\
${3/2}^{-}_{1}$ & 51 & 31 & 5 & 11 & 2 \\
${3/2}^{-}_{2}$ & 0 & 28 & 57 & 7 & 8 \\
${3/2}^{-}_{3}$ & 16 & 40 & 31 & 7 & 6 \\
${5/2}^{-}_{1}$ & 53 & 6 & 30 & 3 & 8 \\
${5/2}^{-}_{2}$ & 15 & 22 & 50 & 6 & 7 \\
${5/2}^{-}_{3}$ & 3 & 51 & 30 & 11 & 5 \\
${7/2}^{-}_{1}$ & 44 & 34 & 7 & 13 & 2 \\
${7/2}^{-}_{2}$ & 4 & 16 & 66 & 4 & 10 \\
${7/2}^{-}_{3}$ & 19 & 44 & 23 & 10 & 4 \\
${9/2}^{-}_{1}$ & 47 & 5 & 35 & 2 & 11 \\
${9/2}^{-}_{2}$ & 19 & 19 & 48 & 4 & 9 \\
${9/2}^{-}_{3}$ & 2 & 49 & 32 & 12 & 4 \\
${11/2}^{-}_{1}$ & 42 & 34 & 8 & 14 & 2 \\
${11/2}^{-}_{2}$ & 3 & 14 & 68 & 3 & 12 \\
${13/2}^{-}_{1}$ & 45 & 5 & 37 & 2 & 11 \\
\hline\hline
\end{tabular*}
\end{center}
\end{table}


\begin{figure}[htb!]
\begin{center}
\includegraphics[width=\linewidth]{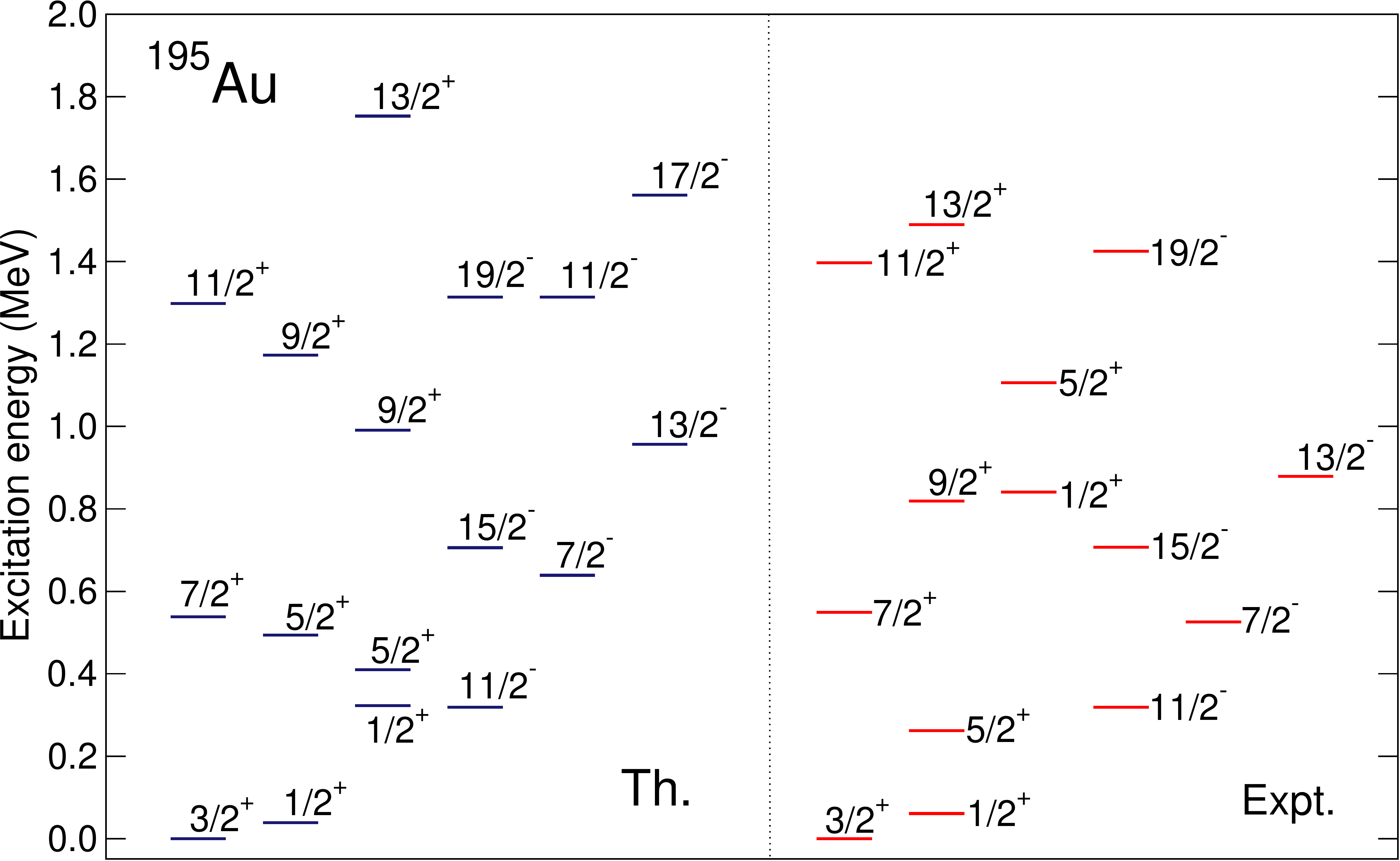}
\caption{(Color online) Same as in  Fig.~\ref{fig:195pt}, but for 
 $^{195}$Au.}
\label{fig:195au}
\end{center}
\end{figure}

\begin{table}[hb!]
\caption{\label{tab:195au-frac} 
Amplitudes (in per cent) of the  
positive-parity states of $^{195}$Au shown in Fig.~\ref{fig:195au}
when expressed in the single particle basis of the $3s_{1/2}$,
$2d_{3/2}$, $2d_{5/2}$ and $1g_{7/2}$ orbitals.}
\begin{center}
\begin{tabular*}{\columnwidth}{p{1.6cm}p{1.6cm}p{1.6cm}p{1.6cm}p{1.6cm}}
\hline\hline
\textrm{$J^{\pi}$} &
\textrm{$3s_{1/2}$}&
\textrm{$2d_{3/2}$}&
\textrm{$2d_{5/2}$}&
\textrm{$1g_{7/2}$} \\
\hline
${1/2}^{+}_{1}$ & 59 & 27 & 12 & 2 \\
${1/2}^{+}_{2}$ & 14 & 61 & 15 & 10 \\
${3/2}^{+}_{1}$ & 3 & 84 & 2 & 11 \\
${5/2}^{+}_{1}$ & 1 & 83 & 6 & 10 \\
${5/2}^{+}_{2}$ & 67 & 11 & 21 & 1 \\
${7/2}^{+}_{1}$ & 2 & 83 & 1 & 14 \\
${9/2}^{+}_{1}$ & 8 & 73 & 9 & 10 \\
${9/2}^{+}_{2}$ & 60 & 19 & 20 & 1 \\
${11/2}^{+}_{1}$ & 1 & 83 & 2 & 14 \\
${13/2}^{+}_{1}$ & 14 & 68 & 9 & 9 \\
\hline\hline
\end{tabular*}
\end{center}
\end{table}


\begin{table}[htb]
\caption{\label{tab:195pt}%
$B(E2)$ and $B(M1)$ transition probabilities  (in Weisskopf units) for $^{195}$Pt. }
\begin{center}
\begin{tabular}{p{2.5cm}cccc}
\hline\hline
\multirow{2}{*}{} & \multicolumn{2}{c}{$B(E2)$ (W.u.)} &
 \multicolumn{2}{c}{$B(M1)$ (W.u.)} \\
\cline{2-3} 
\cline{4-5}
          & Th.         & Exp.    & Th.         & Exp.     \\
\hline
${3/2}^-_1\rightarrow {1/2}^-_1$ & 36 & 11.5(15) & 3.9$\times 10^{-5}$ & 0.0168(19) \\
${3/2}^-_2\rightarrow {1/2}^-_1$ & 0.086 & 4.5(13) & 0.00032 & 0.00033(11) \\
${3/2}^-_3\rightarrow {1/2}^-_1$ & 6.1 & 30(7) & 0.023 & 0.024(4) \\
${3/2}^-_4\rightarrow {1/2}^-_1$ & 2.9 & 0.22(7) & 0.011 & 0.0036(7) \\
${3/2}^-_4\rightarrow {1/2}^-_2$ & 4.2 & $<$37 & 0.016 & $>$0.00054 \\
${5/2}^-_1\rightarrow {1/2}^-_1$ & 35 & 8.9(7) & - & - \\
${5/2}^-_2\rightarrow {1/2}^-_1$ & 7.5 & 49(7) & - & - \\
${5/2}^-_3\rightarrow {1/2}^-_1$ & 0.0093 & 1.3(9) & - & - \\
${3/2}^-_2\rightarrow {3/2}^-_1$ & 4.0 & 0.05$^{+106}_{-5}$ & 0.0043 & 0.0030(8) \\
${3/2}^-_4\rightarrow {3/2}^-_1$ & 6.8 & 0.07(6) & 4.3$\times 10^{-5}$  & 0.0013(3) \\
${5/2}^-_1\rightarrow {3/2}^-_1$ & 9.9 & 4.8(19) & 0.017 & 0.0269(21) \\
${5/2}^-_2\rightarrow {3/2}^-_1$ & 0.076 & 11(6) & 9.9$\times 10^{-5}$ & 0.019(3) \\
${5/2}^-_3\rightarrow {3/2}^-_1$ & 7.2 & 38(20) & 0.027 & 0.038(17) \\
${5/2}^-_4\rightarrow {3/2}^-_1$ & - & - & 0.0097 & $<0.013$ \\
${5/2}^-_4\rightarrow {3/2}^-_3$ & - & - & 0.0030 & $<0.017$\\
${7/2}^-_2\rightarrow {3/2}^-_1$ & 0.84 & 29(10) & - & - \\
${7/2}^-_2\rightarrow {3/2}^-_3$ & 4.0 & 7(3) & - & - \\
${7/2}^-_3\rightarrow {3/2}^-_3$ & 34 & 26(17) & - & - \\
${5/2}^-_3\rightarrow {5/2}^-_1$ & 1.9 & 0.015$^{+88}_{-15}$ & 0.0044 &
 0.026(12)\\
${7/2}^-_2\rightarrow {5/2}^-_1$ & - & - & 0.033 & 0.014(5) \\
${9/2}^-_1\rightarrow {5/2}^-_1$ & 57 & 35(8) & - & - \\
${5/2}^-_3\rightarrow {5/2}^-_2$ & - & - & 0.027 & 0.030(15) \\
${5/2}^-_4\rightarrow {5/2}^-_2$ & 6.3 & $<60$ & - & - \\
${7/2}^-_3\rightarrow {5/2}^-_2$ & 1.7 & $<210$ & 0.0064 & $<0.077$ \\
${9/2}^-_2\rightarrow {5/2}^-_2$ & 47 & 30(8) & - & - \\
${7/2}^-_2\rightarrow {5/2}^-_3$ & 2.1 & $<3.9\times 10^{3}$ & 0.0080 & $<0.14$ \\
\hline\hline
\end{tabular}
\end{center}
\end{table}

In this section we turn our attention to  the $\gamma$-soft cases. In 
Figs.~\ref{fig:195pt} and \ref{fig:195au} we have plotted the three 
lowest positive- and negative-parity bands obtained for both $^{195}$Pt 
and $^{195}$Au. They are compared with the experimental ones taken from Ref. 
\cite{data}. 
To understand the structure of those states, we show in 
Tables~\ref{tab:195pt-frac} and \ref{tab:195au-frac} the amplitudes of 
their decomposition in the spherical single particle basis. The two 
tables correspond to the results for $^{195}$Pt and $^{195}$Au, 
respectively.
It should be noted that, in Figs.~\ref{fig:195pt} and
\ref{fig:195au}, the theoretical levels have been classified into bands
according to the dominant E2 decays and also that the relative location
of the ground states for positive and negative parity in the theoretical
energy spectrum has been adjusted to that of the experimental data.

As can be seen from Fig.~\ref{fig:195pt}, our calculations reproduce 
fairly well the  experimental energies of the negative-parity states in 
$^{195}$Pt, whereas many of the non-yrast states are overestimated. 
Both theoretically and experimentally, the lowest negative-parity band 
appears to show the $\Delta J=1$ systematics of the strong coupling 
limit though, there is a staggering pattern (${3/2}^-,{5/2}^-$), 
(${7/2}^-,{9/2}^-$), etc. Consistent with the experiment, the first and 
second excited negative-parity bands also display a $\Delta J=1$ 
feature. Nevertheless, their bandhead energies  are rather high in 
comparison with the experimental ones. Furthermore, we have classified 
the calculated ${3/2}^-_1$ and ${5/2}^-_{1}$ levels, which are nearly 
degenerated, into the lowest band, whereas the experimental ${3/2}^-_1$ 
and ${5/2}^-_{1}$ levels are suggested to be the bandheads of the first 
and second excited bands, respectively. As seen from
Table ~\ref{tab:195pt-frac}. the features already mentioned 
can be understood from the fact that, in our calculations, the states 
in the lowest negative-parity band are predominantly composed of the 
$p_{1/2}$ configuration while those in the first and second excited 
negative-parity bands are mainly composed  of the $f_{5/2}$ and 
$p_{3/2}$ configurations, respectively. 
From Table~\ref{tab:spe} we note that the $3p_{3/2}$ and $2f_{5/2}$ single-particle
levels lie much higher than the $3p_{1/2}$ orbital. This could partly
account for the discrepancy observed in $^{195}$Pt where the first and 
second excited negative-parity bands are predicted to lie too high in
excitation energy as compared to the experiment. 
On the other hand, the pattern 
of  the positive-parity levels in $^{195}$Pt, also shown in 
Fig.~\ref{fig:195pt}, is much simpler than for the negative-parity 
ones. The ground state for positive parity, the ${13/2}^+_1$ state, is weakly
coupled to the boson core nucleus $^{194}$Pt. Consistent with the
experiment, the three theoretical positive-parity bands, shown in the
left-hand side of Fig.~\ref{fig:195pt}, look rather
harmonic. They exhibit the weak coupling $\Delta J=2$ systematics.

The results obtained for $^{195}$Au are shown in Fig.~\ref{fig:195au}. 
In this case, our calculations provide a slightly better agreement with 
the experimental data  than for $^{195}$Pt. The three lowest 
positive-parity bands display the $\Delta J=2$ systematics of the weak 
coupling limit. This is consistent with the experimental data. 
Nevertheless, the calculated first excited positive-parity band looks 
more stretched than the experimental one. Also, the second excited 
positive-parity band, built on the ${1/2}^+_2$ state, is much lower in 
energy in the present calculation than in experiment. In our 
calculation, the low-lying positive-parity states of the $^{195}$Au 
nucleus are mainly composed of the $s_{1/2}$ and $d_{3/2}$ 
configurations. For example, 84 \% of the ground state ${3/2}^+_1$ is 
made of the $d_{3/2}$ configuration, while 59 \% of the first excited 
state ${1/2}^+_1$ is comprised of the $s_{1/2}$ configuration (see,
Table ~\ref{tab:195au-frac}). 
Similarly to the positive-parity bands in $^{195}$Pt nucleus, the
theoretical negative-parity bands in $^{195}$Au all exhibit the
weak-coupling $\Delta J=2$ systematics and look harmonic. 


\begin{table}[htb]
\caption{\label{tab:195pt_mom}
Spectroscopic quadrupole (in $eb$) and magnetic moments (in $\mu_N^2$) 
in $^{195}$Pt.}
\begin{center}
\begin{tabular}{p{2.5cm}cccc}
\hline\hline
\multirow{2}{*}{} & \multicolumn{2}{c}{$Q_J$ ($e$b)} &
 \multicolumn{2}{c}{$\mu_J$ ($\mu_N^2$)} \\
\cline{2-3} 
\cline{4-5}
          & Th.         & Exp.    & Th.         & Exp.     \\
\hline
${13/2}^+_1$ & +1.619 & +1.4(6) & -1.246 & -0.606(105) \\
${1/2}^-_1$ & - & - & +0.467 & +0.60952(6) \\
${3/2}^-_1$ & +0.683 & - & -0.206 & -0.62(6) \\
${3/2}^-_3$ & +0.288 & - & -0.411 & +0.16(3) \\
${5/2}^-_1$ & +0.973 & - & +0.924 & +0.90(6) \\
${5/2}^-_2$ & +0.306 & - & +1.219 & +0.52(5) \\
${5/2}^-_3$ & +0.536 & - & +0.201 & +0.39(10) \\
${5/2}^-_4$ & -0.729 & - & +0.925 & +1.6(6) \\
${7/2}^-_2$ & +0.637 & - & +1.079 & +0.55(8) \\
${7/2}^-_3$ & +0.566 & - & +0.270 & +1.4(4) \\
${7/2}^-_5$ & +0.179 & - & +0.212 & +1.2(3) \\
${9/2}^-_2$ & +0.731 & - & +1.735 & +1.55(12) \\
${9/2}^-_3$ & +0.744 & - & +0.785 & +1.52(16) \\
\hline\hline
\end{tabular}
\end{center}
\end{table}


The calculated $B(E2)$ and $B(M1)$ transition rates of the $^{195}$Pt 
nucleus are shown in Table~\ref{tab:195pt} and compared with the 
experimental data of Ref. \cite{data,stone2005}. As it is apparent from 
the table, the agreement between our calculation and experiment is not 
necessarily satisfactory in some of the $B(E2)$ transition rates. For 
instance, the theoretical $B(E2; {5/2}^-_2\rightarrow {3/2}^-_1)$ value 
of 0.076 W.u. is roughly a factor 10$^2$ smaller than the experimental 
value of 11$\pm$6 W.u. A possible reason for the disagreement could be 
attributed to the choice of both the boson and fermion effective 
charges. A more likely reason would be the fact that the structures of 
the ${5/2}^-_2$ and ${3/2}^-_1$ wave functions are somewhat different 
in the present calculation (see, Table~\ref{tab:195pt-frac}). The
${3/2}^-_1$ state is mostly composed of the $p_{1/2}$ (51 \%) and
$p_{3/2}$ (31 \%) configurations. On the other hand, the main component
of the ${5/2}^-_2$ wave function is the $f_{5/2}$ configuration (50 \%),
while the $p_{1/2}$ and $p_{3/2}$ configurations account for 15 \% and
22 \% of the wave function, respectively. All in all, the predicted
$B(M1)$ values compare well the experimental data. We also compare in
Table~\ref{tab:195pt_mom} the predicted and experimental
\cite{data,stone2005} spectroscopic 
quadrupole $Q_{J}$ and magnetic $\mu_J$ moments of the $^{195}$Pt 
nucleus. In this table, the calculated $Q_{{13/2}^+_1}$ value is in 
good agreement with the experimental one. We also show in 
Table~\ref{tab:195pt_mom} the predicted $Q_J$ values for other states, 
where data are not available. 
The sign of the predicted $\mu_J$ values is, in most cases, consistent with the data, apart from 
$\mu_{{3/2}^-_3}$.


\begin{table}[htb]
\caption{\label{tab:195au}%
$B(E2)$ and $B(M1)$ transition strengths, and spectroscopic
 quadrupole and magnetic moments in $^{195}$Au. }
\begin{center}
\begin{tabular}{p{2.5cm}cccc}
\hline\hline
\multirow{2}{*}{} & \multicolumn{2}{c}{$B(E2)$ (W.u.)} &
 \multicolumn{2}{c}{$B(M1)$ (W.u.)} \\
\cline{2-3} 
\cline{4-5}
          & Th.         & Exp.    & Th.         & Exp.     \\
\hline
${1/2}^+_1\rightarrow {3/2}^+_1$ & 31 & 41(4) & 0.102 & 0.00199(15) \\
${3/2}^+_2\rightarrow {1/2}^+_1$ & 31 & $>15$ & 0.114 & $>0.051$ \\
${3/2}^+_2\rightarrow {3/2}^+_1$ & 1.4 & $>4.3$ & 0.0062 & $>0.00013$ \\
${5/2}^+_1\rightarrow {3/2}^+_2$ & 0.62 & 8.7(20) & - & - \\
${5/2}^+_1\rightarrow {3/2}^+_1$ & 39 & 18(4) & 9.3$\times 10^{-5}$ & 0.0124(25) \\
${25/2}^+_1\rightarrow {21/2}^+_1$ & 41 & 10.9(25) & - & - \\
\hline
\multirow{2}{*}{} & \multicolumn{2}{c}{$Q_J$ ($e$b)} &
 \multicolumn{2}{c}{$\mu_J$ ($\mu_N^2$)} \\
\cline{2-3} 
\cline{4-5}
          & Th.         & Exp.    & Th.         & Exp.     \\
\hline
${3/2}^+_1$ & +0.781 & +0.607(18) & +0.633 & +0.145(5) \\
${11/2}^-_1$ & +1.708 & +1.87(6) & +6.70 & +6.17(9) \\
\hline\hline
\end{tabular}
\end{center}
\end{table}

In Table~\ref{tab:195au} we compare the calculated $B(E2)$, $B(M1)$, 
$Q_J$ and $\mu_J$ values of the $^{195}$Au nucleus with the 
experimental data. Overall, the calculation reproduces experimental 
data rather well. As mentioned above, some disagreement between the 
calculated and experimental $B(E2)$ values could partly arise from the 
chosen effective charges for the fermion and boson quadrupole operators 
although it is more likely to be due to the differences in the nature 
of the wave functions of the initial and final states. In the case of 
the $B(E2; {5/2}^+_1\rightarrow {3/2}^+_2)$ transition rate, for 
instance, the wave function of the ${5/2}^+_1$ state in $^{195}$Au is 
made predominantly of the $d_{3/2}$ single-particle configuration (83 
\%), while the ${3/2}^+_2$ wave function is mainly composed of the 
$s_{1/2}$ configuration (66 \%). For more details, see
Table~\ref{tab:195au-frac}. One realizes in Table~\ref{tab:195au} that
the calculated $Q_J$ and $\mu_J$ values for the lowest positive-
(${3/2}^+_1$) and negative-parity (${11/2}^-_1$) states of the
$^{195}$Au nucleus are in an excellent agreement with the available
experimental data \cite{data,stone2005}.


\section{Summary and concluding remarks\label{sec:summary}}


In this work we have studied the spectroscopic properties of several 
odd-mass nuclei within the IBFM framework based on the Gogny-D1M EDF. 
Following the procedure developed in Ref.~\cite{nomura2016odd}, the 
($\beta,\gamma$)-deformation energy surface for the even-even core 
nuclei, as well as the single-particle energies and occupation 
probabilities of the odd nucleon, have been provided by the 
self-consistent HFB method based on the Gogny-D1M EDF, and have been 
used as a microscopic input for the construction of the Hamiltonian of 
the IBFM. As was done in the original work of 
Ref.~\cite{nomura2016odd}, the three strength parameters of the 
particle-core coupling Hamiltonian have been fitted in each of the 
odd-mass nucleus considered as to reproduce selected experimental data 
for the low-energy excitation spectra. The method has been applied to 
the axially-deformed odd-mass $^{149-155}$Eu and $^{149-155}$Sm nuclei, 
and to the $\gamma$-soft odd-mass $^{195}$Pt and $^{195}$Au nuclei. 

The present calculation describes fairly well the experimental 
systematics of excitation spectra and electromagnetic properties of the 
odd-mass Eu and Sm nuclei as signatures of structural evolution from 
the nearly spherical to axially-deformed shapes. Our calculation on the 
odd-mass Eu and Sm nuclei reveals the same level of accuracy in 
describing the odd-mass isotopes as the previous calculations on the 
same nuclei \cite{nomura2016odd,nomura2016qpt} based on a relativistic 
EDF. The method also provides a reasonable description of the 
low-energy excitation spectra in the $\gamma$-soft $^{195}$Pt and 
$^{195}$Au nuclei, whereas the agreement between the calculated and 
experimental electromagnetic properties is not entirely satisfactory. 
Nevertheless, considering the fact the model involves only  three 
phenomenological parameters, all these results for the $\gamma$-soft 
systems seem to be rather promising. 

Our next step would be to apply the method to study more systematically 
the structural evolution in other $\gamma$-soft odd-mass systems, 
either because they are supposed to have a rich spectrum and/or are of 
experimental interest. A more ambitious perspective is the 
determination of the boson-fermion coupling Hamiltonian parameters from 
quantities obtained solely from the mean field calculations with the 
Gogny EDF. This possibility will give us the key to make true 
predictions for odd-A systems where no experimental data exists. Work 
along these lines are in progress and will be reported elsewhere.

\acknowledgments
We would like to thank D. Vretenar and T. Nik\v si\'c for useful discussions. 
K.N. acknowledges support from the Japan 
Society for the Promotion of Science. This work has been supported in 
part by the QuantiXLie Centre of Excellence. The  work of LMR was 
supported by Spanish grant Nos FPA2015-65929-P MINECO and
FIS2015-63770-P MINECO.

\bibliography{refs}

\begin{thebibliography}{41}%
\makeatletter
\providecommand \@ifxundefined [1]{%
 \@ifx{#1\undefined}
}%
\providecommand \@ifnum [1]{%
 \ifnum #1\expandafter \@firstoftwo
 \else \expandafter \@secondoftwo
 \fi
}%
\providecommand \@ifx [1]{%
 \ifx #1\expandafter \@firstoftwo
 \else \expandafter \@secondoftwo
 \fi
}%
\providecommand \natexlab [1]{#1}%
\providecommand \enquote  [1]{``#1''}%
\providecommand \bibnamefont  [1]{#1}%
\providecommand \bibfnamefont [1]{#1}%
\providecommand \citenamefont [1]{#1}%
\providecommand \href@noop [0]{\@secondoftwo}%
\providecommand \href [0]{\begingroup \@sanitize@url \@href}%
\providecommand \@href[1]{\@@startlink{#1}\@@href}%
\providecommand \@@href[1]{\endgroup#1\@@endlink}%
\providecommand \@sanitize@url [0]{\catcode `\\12\catcode `\$12\catcode
  `\&12\catcode `\#12\catcode `\^12\catcode `\_12\catcode `\%12\relax}%
\providecommand \@@startlink[1]{}%
\providecommand \@@endlink[0]{}%
\providecommand \url  [0]{\begingroup\@sanitize@url \@url }%
\providecommand \@url [1]{\endgroup\@href {#1}{\urlprefix }}%
\providecommand \urlprefix  [0]{URL }%
\providecommand \Eprint [0]{\href }%
\providecommand \doibase [0]{http://dx.doi.org/}%
\providecommand \selectlanguage [0]{\@gobble}%
\providecommand \bibinfo  [0]{\@secondoftwo}%
\providecommand \bibfield  [0]{\@secondoftwo}%
\providecommand \translation [1]{[#1]}%
\providecommand \BibitemOpen [0]{}%
\providecommand \bibitemStop [0]{}%
\providecommand \bibitemNoStop [0]{.\EOS\space}%
\providecommand \EOS [0]{\spacefactor3000\relax}%
\providecommand \BibitemShut  [1]{\csname bibitem#1\endcsname}%
\let\auto@bib@innerbib\@empty
\bibitem [{\citenamefont {Bohr}\ and\ \citenamefont {Mottelsson}(1975)}]{BM}%
  \BibitemOpen
  \bibfield  {author} {\bibinfo {author} {\bibfnamefont {A.}~\bibnamefont
  {Bohr}}\ and\ \bibinfo {author} {\bibfnamefont {B.~M.}\ \bibnamefont
  {Mottelsson}},\ }\href@noop {} {\emph {\bibinfo {title} {Nuclear
  Structure}}},\ Vol.~\bibinfo {volume} {2}\ (\bibinfo  {publisher} {Benjamin,
  New York, USA},\ \bibinfo {year} {1975})\ p.~\bibinfo {pages}
  {45}\BibitemShut {NoStop}%
\bibitem [{\citenamefont {Bohr}(1953)}]{bohr1953}%
  \BibitemOpen
  \bibfield  {author} {\bibinfo {author} {\bibfnamefont {A.}~\bibnamefont
  {Bohr}},\ }\href@noop {} {\bibfield  {journal} {\bibinfo  {journal} {Mat.
  Fys. Medd. Dan. Vid. Selsk.}\ }\textbf {\bibinfo {volume} {27}},\ \bibinfo
  {pages} {16} (\bibinfo {year} {1953})}\BibitemShut {NoStop}%
\bibitem [{\citenamefont {Caurier}\ \emph {et~al.}(2005)\citenamefont
  {Caurier}, \citenamefont {Mart\'{\i}nez-Pinedo}, \citenamefont {Nowacki},
  \citenamefont {Poves},\ and\ \citenamefont {Zuker}}]{caurier2005}%
  \BibitemOpen
  \bibfield  {author} {\bibinfo {author} {\bibfnamefont {E.}~\bibnamefont
  {Caurier}}, \bibinfo {author} {\bibfnamefont {G.}~\bibnamefont
  {Mart\'{\i}nez-Pinedo}}, \bibinfo {author} {\bibfnamefont {F.}~\bibnamefont
  {Nowacki}}, \bibinfo {author} {\bibfnamefont {A.}~\bibnamefont {Poves}}, \
  and\ \bibinfo {author} {\bibfnamefont {A.~P.}\ \bibnamefont {Zuker}},\
  }\href@noop {} {\bibfield  {journal} {\bibinfo  {journal} {Rev. Mod. Phys.}\
  }\textbf {\bibinfo {volume} {77}},\ \bibinfo {pages} {427} (\bibinfo {year}
  {2005})}\BibitemShut {NoStop}%
\bibitem [{\citenamefont {Bender}\ \emph {et~al.}(2003)\citenamefont {Bender},
  \citenamefont {Heenen},\ and\ \citenamefont {Reinhard}}]{bender2003}%
  \BibitemOpen
  \bibfield  {author} {\bibinfo {author} {\bibfnamefont {M.}~\bibnamefont
  {Bender}}, \bibinfo {author} {\bibfnamefont {P.-H.}\ \bibnamefont {Heenen}},
  \ and\ \bibinfo {author} {\bibfnamefont {P.-G.}\ \bibnamefont {Reinhard}},\
  }\href {\doibase 10.1103/RevModPhys.75.121} {\bibfield  {journal} {\bibinfo
  {journal} {Rev. Mod. Phys.}\ }\textbf {\bibinfo {volume} {75}},\ \bibinfo
  {pages} {121} (\bibinfo {year} {2003})}\BibitemShut {NoStop}%
\bibitem [{\citenamefont {Robledo}\ \emph {et~al.}(2012)\citenamefont
  {Robledo}, \citenamefont {Bernard},\ and\ \citenamefont
  {Bertsch}}]{PhysRevC.86.064313}%
  \BibitemOpen
  \bibfield  {author} {\bibinfo {author} {\bibfnamefont {L.~M.}\ \bibnamefont
  {Robledo}}, \bibinfo {author} {\bibfnamefont {R.}~\bibnamefont {Bernard}}, \
  and\ \bibinfo {author} {\bibfnamefont {G.~F.}\ \bibnamefont {Bertsch}},\
  }\href {\doibase 10.1103/PhysRevC.86.064313} {\bibfield  {journal} {\bibinfo
  {journal} {Phys. Rev. C}\ }\textbf {\bibinfo {volume} {86}},\ \bibinfo
  {pages} {064313} (\bibinfo {year} {2012})}\BibitemShut {NoStop}%
\bibitem [{\citenamefont {Rodr\'iguez-Guzm\'an}\ \emph
  {et~al.}(2010{\natexlab{a}})\citenamefont {Rodr\'iguez-Guzm\'an},
  \citenamefont {Sarriguren}, \citenamefont {Robledo},\ and\ \citenamefont
  {Perez-Martin}}]{rayner10odd-1}%
  \BibitemOpen
  \bibfield  {author} {\bibinfo {author} {\bibfnamefont {R.}~\bibnamefont
  {Rodr\'iguez-Guzm\'an}}, \bibinfo {author} {\bibfnamefont {P.}~\bibnamefont
  {Sarriguren}}, \bibinfo {author} {\bibfnamefont {L.~M.}\ \bibnamefont
  {Robledo}}, \ and\ \bibinfo {author} {\bibfnamefont {S.}~\bibnamefont
  {Perez-Martin}},\ }\href {\doibase 10.1016/j.physletb.2010.06.035} {\bibfield
   {journal} {\bibinfo  {journal} {Phys. Lett. B}\ }\textbf {\bibinfo {volume}
  {691}},\ \bibinfo {pages} {202 } (\bibinfo {year}
  {2010}{\natexlab{a}})}\BibitemShut {NoStop}%
\bibitem [{\citenamefont {Rodr\'iguez-Guzm\'an}\ \emph
  {et~al.}(2010{\natexlab{b}})\citenamefont {Rodr\'iguez-Guzm\'an},
  \citenamefont {Sarriguren},\ and\ \citenamefont {Robledo}}]{rayner10odd-2}%
  \BibitemOpen
  \bibfield  {author} {\bibinfo {author} {\bibfnamefont {R.}~\bibnamefont
  {Rodr\'iguez-Guzm\'an}}, \bibinfo {author} {\bibfnamefont {P.}~\bibnamefont
  {Sarriguren}}, \ and\ \bibinfo {author} {\bibfnamefont {L.~M.}\ \bibnamefont
  {Robledo}},\ }\href {\doibase 10.1103/PhysRevC.82.044318} {\bibfield
  {journal} {\bibinfo  {journal} {Phys. Rev. C}\ }\textbf {\bibinfo {volume}
  {82}},\ \bibinfo {pages} {044318} (\bibinfo {year}
  {2010}{\natexlab{b}})}\BibitemShut {NoStop}%
\bibitem [{\citenamefont {Rodr\'iguez-Guzm\'an}\ \emph
  {et~al.}(2010{\natexlab{c}})\citenamefont {Rodr\'iguez-Guzm\'an},
  \citenamefont {Sarriguren},\ and\ \citenamefont {Robledo}}]{rayner10odd-3}%
  \BibitemOpen
  \bibfield  {author} {\bibinfo {author} {\bibfnamefont {R.}~\bibnamefont
  {Rodr\'iguez-Guzm\'an}}, \bibinfo {author} {\bibfnamefont {P.}~\bibnamefont
  {Sarriguren}}, \ and\ \bibinfo {author} {\bibfnamefont {L.~M.}\ \bibnamefont
  {Robledo}},\ }\href {\doibase 10.1103/PhysRevC.82.061302} {\bibfield
  {journal} {\bibinfo  {journal} {Phys. Rev. C}\ }\textbf {\bibinfo {volume}
  {82}},\ \bibinfo {pages} {061302} (\bibinfo {year}
  {2010}{\natexlab{c}})}\BibitemShut {NoStop}%
\bibitem [{\citenamefont {Bally}\ \emph {et~al.}(2014)\citenamefont {Bally},
  \citenamefont {Avez}, \citenamefont {Bender},\ and\ \citenamefont
  {Heenen}}]{bally2014}%
  \BibitemOpen
  \bibfield  {author} {\bibinfo {author} {\bibfnamefont {B.}~\bibnamefont
  {Bally}}, \bibinfo {author} {\bibfnamefont {B.}~\bibnamefont {Avez}},
  \bibinfo {author} {\bibfnamefont {M.}~\bibnamefont {Bender}}, \ and\ \bibinfo
  {author} {\bibfnamefont {P.-H.}\ \bibnamefont {Heenen}},\ }\href@noop {}
  {\bibfield  {journal} {\bibinfo  {journal} {Phys. Rev. Lett.}\ }\textbf
  {\bibinfo {volume} {113}},\ \bibinfo {pages} {162501} (\bibinfo {year}
  {2014})}\BibitemShut {NoStop}%
\bibitem [{\citenamefont {Borrajo}\ and\ \citenamefont
  {Egido}(2016)}]{Borrajo2016}%
  \BibitemOpen
  \bibfield  {author} {\bibinfo {author} {\bibfnamefont {M.}~\bibnamefont
  {Borrajo}}\ and\ \bibinfo {author} {\bibfnamefont {J.~L.}\ \bibnamefont
  {Egido}},\ }\href {\doibase 10.1140/epja/i2016-16277-8} {\bibfield  {journal}
  {\bibinfo  {journal} {The European Physical Journal A}\ }\textbf {\bibinfo
  {volume} {52}},\ \bibinfo {pages} {277} (\bibinfo {year} {2016})}\BibitemShut
  {NoStop}%
\bibitem [{\citenamefont {Litvinova}\ and\ \citenamefont
  {Afanasjev}(2011)}]{litvinova2011}%
  \BibitemOpen
  \bibfield  {author} {\bibinfo {author} {\bibfnamefont {E.~V.}\ \bibnamefont
  {Litvinova}}\ and\ \bibinfo {author} {\bibfnamefont {A.~V.}\ \bibnamefont
  {Afanasjev}},\ }\href@noop {} {\bibfield  {journal} {\bibinfo  {journal}
  {Phys. Rev. C}\ }\textbf {\bibinfo {volume} {84}},\ \bibinfo {pages} {014305}
  (\bibinfo {year} {2011})}\BibitemShut {NoStop}%
\bibitem [{\citenamefont {Mizuyama}\ \emph {et~al.}(2012)\citenamefont
  {Mizuyama}, \citenamefont {Col\`o},\ and\ \citenamefont
  {Vigezzi}}]{mizuyama2012}%
  \BibitemOpen
  \bibfield  {author} {\bibinfo {author} {\bibfnamefont {K.}~\bibnamefont
  {Mizuyama}}, \bibinfo {author} {\bibfnamefont {G.}~\bibnamefont {Col\`o}}, \
  and\ \bibinfo {author} {\bibfnamefont {E.}~\bibnamefont {Vigezzi}},\
  }\href@noop {} {\bibfield  {journal} {\bibinfo  {journal} {Phys. Rev. C}\
  }\textbf {\bibinfo {volume} {86}},\ \bibinfo {pages} {034318} (\bibinfo
  {year} {2012})}\BibitemShut {NoStop}%
\bibitem [{\citenamefont {Niu}\ \emph {et~al.}(2015)\citenamefont {Niu},
  \citenamefont {Niu}, \citenamefont {Col\`o},\ and\ \citenamefont
  {Vigezzi}}]{niu2015}%
  \BibitemOpen
  \bibfield  {author} {\bibinfo {author} {\bibfnamefont {Y.~F.}\ \bibnamefont
  {Niu}}, \bibinfo {author} {\bibfnamefont {Z.~M.}\ \bibnamefont {Niu}},
  \bibinfo {author} {\bibfnamefont {G.}~\bibnamefont {Col\`o}}, \ and\ \bibinfo
  {author} {\bibfnamefont {E.}~\bibnamefont {Vigezzi}},\ }\href@noop {}
  {\bibfield  {journal} {\bibinfo  {journal} {Phys. Rev. Lett.}\ }\textbf
  {\bibinfo {volume} {114}},\ \bibinfo {pages} {142501} (\bibinfo {year}
  {2015})}\BibitemShut {NoStop}%
\bibitem [{\citenamefont {De~Gregorio}\ \emph {et~al.}(2016)\citenamefont
  {De~Gregorio}, \citenamefont {Knapp}, \citenamefont {Lo~Iudice},\ and\
  \citenamefont {Vesely}}]{degregorio2016}%
  \BibitemOpen
  \bibfield  {author} {\bibinfo {author} {\bibfnamefont {G.}~\bibnamefont
  {De~Gregorio}}, \bibinfo {author} {\bibfnamefont {F.}~\bibnamefont {Knapp}},
  \bibinfo {author} {\bibfnamefont {N.}~\bibnamefont {Lo~Iudice}}, \ and\
  \bibinfo {author} {\bibfnamefont {P.}~\bibnamefont {Vesely}},\ }\href
  {\doibase 10.1103/PhysRevC.94.061301} {\bibfield  {journal} {\bibinfo
  {journal} {Phys. Rev. C}\ }\textbf {\bibinfo {volume} {94}},\ \bibinfo
  {pages} {061301} (\bibinfo {year} {2016})}\BibitemShut {NoStop}%
\bibitem [{\citenamefont {De~Gregorio}\ \emph {et~al.}(2017)\citenamefont
  {De~Gregorio}, \citenamefont {Knapp}, \citenamefont {Lo~Iudice},\ and\
  \citenamefont {Vesel\'y}}]{degregorio2017}%
  \BibitemOpen
  \bibfield  {author} {\bibinfo {author} {\bibfnamefont {G.}~\bibnamefont
  {De~Gregorio}}, \bibinfo {author} {\bibfnamefont {F.}~\bibnamefont {Knapp}},
  \bibinfo {author} {\bibfnamefont {N.}~\bibnamefont {Lo~Iudice}}, \ and\
  \bibinfo {author} {\bibfnamefont {P.}~\bibnamefont {Vesel\'y}},\ }\href
  {\doibase 10.1103/PhysRevC.95.034327} {\bibfield  {journal} {\bibinfo
  {journal} {Phys. Rev. C}\ }\textbf {\bibinfo {volume} {95}},\ \bibinfo
  {pages} {034327} (\bibinfo {year} {2017})}\BibitemShut {NoStop}%
\bibitem [{\citenamefont {Scholten}(1985)}]{scholten1985}%
  \BibitemOpen
  \bibfield  {author} {\bibinfo {author} {\bibfnamefont {O.}~\bibnamefont
  {Scholten}},\ }\href@noop {} {\bibfield  {journal} {\bibinfo  {journal}
  {Prog. Part. Nucl. Phys.}\ }\textbf {\bibinfo {volume} {14}},\ \bibinfo
  {pages} {189} (\bibinfo {year} {1985})}\BibitemShut {NoStop}%
\bibitem [{\citenamefont {Iachello}\ and\ \citenamefont {{Van
  Isacker}}(1991)}]{IBFM}%
  \BibitemOpen
  \bibfield  {author} {\bibinfo {author} {\bibfnamefont {F.}~\bibnamefont
  {Iachello}}\ and\ \bibinfo {author} {\bibfnamefont {P.}~\bibnamefont {{Van
  Isacker}}},\ }\href@noop {} {\emph {\bibinfo {title} {The interacting
  boson-fermion model}}}\ (\bibinfo  {publisher} {Cambridge University Press,
  Cambridge},\ \bibinfo {year} {1991})\BibitemShut {NoStop}%
\bibitem [{\citenamefont {Iachello}(1981)}]{IBFM-Book}%
  \BibitemOpen
  \bibinfo {editor} {\bibfnamefont {F.}~\bibnamefont {Iachello}},\ ed.,\
  \href@noop {} {\emph {\bibinfo {title} {Interacting Bose-Fermi Systems in
  Nuclei}}}\ (\bibinfo  {publisher} {Springer},\ \bibinfo {address} {New
  York},\ \bibinfo {year} {1981})\BibitemShut {NoStop}%
\bibitem [{\citenamefont {Iachello}\ \emph {et~al.}(2011)\citenamefont
  {Iachello}, \citenamefont {Leviatan},\ and\ \citenamefont
  {Petrellis}}]{iachello2011}%
  \BibitemOpen
  \bibfield  {author} {\bibinfo {author} {\bibfnamefont {F.}~\bibnamefont
  {Iachello}}, \bibinfo {author} {\bibfnamefont {A.}~\bibnamefont {Leviatan}},
  \ and\ \bibinfo {author} {\bibfnamefont {D.}~\bibnamefont {Petrellis}},\
  }\href@noop {} {\bibfield  {journal} {\bibinfo  {journal} {Phys. Lett. B}\
  }\textbf {\bibinfo {volume} {705}},\ \bibinfo {pages} {379 } (\bibinfo {year}
  {2011})}\BibitemShut {NoStop}%
\bibitem [{\citenamefont {Petrellis}\ \emph {et~al.}(2011)\citenamefont
  {Petrellis}, \citenamefont {Leviatan},\ and\ \citenamefont
  {Iachello}}]{petrellis2011}%
  \BibitemOpen
  \bibfield  {author} {\bibinfo {author} {\bibfnamefont {D.}~\bibnamefont
  {Petrellis}}, \bibinfo {author} {\bibfnamefont {A.}~\bibnamefont {Leviatan}},
  \ and\ \bibinfo {author} {\bibfnamefont {F.}~\bibnamefont {Iachello}},\
  }\href@noop {} {\bibfield  {journal} {\bibinfo  {journal} {Ann. Phys.
  (N.Y.)}\ }\textbf {\bibinfo {volume} {326}},\ \bibinfo {pages} {926 }
  (\bibinfo {year} {2011})}\BibitemShut {NoStop}%
\bibitem [{\citenamefont {Nomura}\ \emph
  {et~al.}(2016{\natexlab{a}})\citenamefont {Nomura}, \citenamefont
  {Nik\ifmmode \check{s}\else \v{s}\fi{}i\ifmmode~\acute{c}\else \'{c}\fi{}},\
  and\ \citenamefont {Vretenar}}]{nomura2016odd}%
  \BibitemOpen
  \bibfield  {author} {\bibinfo {author} {\bibfnamefont {K.}~\bibnamefont
  {Nomura}}, \bibinfo {author} {\bibfnamefont {T.}~\bibnamefont {Nik\ifmmode
  \check{s}\else \v{s}\fi{}i\ifmmode~\acute{c}\else \'{c}\fi{}}}, \ and\
  \bibinfo {author} {\bibfnamefont {D.}~\bibnamefont {Vretenar}},\ }\href@noop
  {} {\bibfield  {journal} {\bibinfo  {journal} {Phys. Rev. C}\ }\textbf
  {\bibinfo {volume} {93}},\ \bibinfo {pages} {054305} (\bibinfo {year}
  {2016}{\natexlab{a}})}\BibitemShut {NoStop}%
\bibitem [{\citenamefont {Nik\ifmmode \check{s}\else
  \v{s}\fi{}i\ifmmode~\acute{c}\else \'{c}\fi{}}\ \emph
  {et~al.}(2008)\citenamefont {Nik\ifmmode \check{s}\else
  \v{s}\fi{}i\ifmmode~\acute{c}\else \'{c}\fi{}}, \citenamefont {Vretenar},\
  and\ \citenamefont {Ring}}]{DDPC1}%
  \BibitemOpen
  \bibfield  {author} {\bibinfo {author} {\bibfnamefont {T.}~\bibnamefont
  {Nik\ifmmode \check{s}\else \v{s}\fi{}i\ifmmode~\acute{c}\else \'{c}\fi{}}},
  \bibinfo {author} {\bibfnamefont {D.}~\bibnamefont {Vretenar}}, \ and\
  \bibinfo {author} {\bibfnamefont {P.}~\bibnamefont {Ring}},\ }\href {\doibase
  10.1103/PhysRevC.78.034318} {\bibfield  {journal} {\bibinfo  {journal} {Phys.
  Rev. C}\ }\textbf {\bibinfo {volume} {78}},\ \bibinfo {pages} {034318}
  (\bibinfo {year} {2008})}\BibitemShut {NoStop}%
\bibitem [{\citenamefont {Nomura}\ \emph
  {et~al.}(2016{\natexlab{b}})\citenamefont {Nomura}, \citenamefont
  {Nik\ifmmode \check{s}\else \v{s}\fi{}i\ifmmode~\acute{c}\else \'{c}\fi{}},\
  and\ \citenamefont {Vretenar}}]{nomura2016qpt}%
  \BibitemOpen
  \bibfield  {author} {\bibinfo {author} {\bibfnamefont {K.}~\bibnamefont
  {Nomura}}, \bibinfo {author} {\bibfnamefont {T.}~\bibnamefont {Nik\ifmmode
  \check{s}\else \v{s}\fi{}i\ifmmode~\acute{c}\else \'{c}\fi{}}}, \ and\
  \bibinfo {author} {\bibfnamefont {D.}~\bibnamefont {Vretenar}},\ }\href
  {\doibase 10.1103/PhysRevC.94.064310} {\bibfield  {journal} {\bibinfo
  {journal} {Phys. Rev. C}\ }\textbf {\bibinfo {volume} {94}},\ \bibinfo
  {pages} {064310} (\bibinfo {year} {2016}{\natexlab{b}})}\BibitemShut
  {NoStop}%
\bibitem [{\citenamefont {Nomura}\ \emph {et~al.}(2017)\citenamefont {Nomura},
  \citenamefont {Nik\ifmmode \check{s}\else \v{s}\fi{}i\ifmmode~\acute{c}\else
  \'{c}\fi{}},\ and\ \citenamefont {Vretenar}}]{nomura2017gsoft}%
  \BibitemOpen
  \bibfield  {author} {\bibinfo {author} {\bibfnamefont {K.}~\bibnamefont
  {Nomura}}, \bibinfo {author} {\bibfnamefont {T.}~\bibnamefont {Nik\ifmmode
  \check{s}\else \v{s}\fi{}i\ifmmode~\acute{c}\else \'{c}\fi{}}}, \ and\
  \bibinfo {author} {\bibfnamefont {D.}~\bibnamefont {Vretenar}},\ }\href
  {\doibase 10.1103/PhysRevC.96.014304} {\bibfield  {journal} {\bibinfo
  {journal} {Phys. Rev. C}\ }\textbf {\bibinfo {volume} {96}},\ \bibinfo
  {pages} {014304} (\bibinfo {year} {2017})}\BibitemShut {NoStop}%
\bibitem [{\citenamefont {Goriely}\ \emph {et~al.}(2009)\citenamefont
  {Goriely}, \citenamefont {Hilaire}, \citenamefont {Girod},\ and\
  \citenamefont {P\'eru}}]{D1M}%
  \BibitemOpen
  \bibfield  {author} {\bibinfo {author} {\bibfnamefont {S.}~\bibnamefont
  {Goriely}}, \bibinfo {author} {\bibfnamefont {S.}~\bibnamefont {Hilaire}},
  \bibinfo {author} {\bibfnamefont {M.}~\bibnamefont {Girod}}, \ and\ \bibinfo
  {author} {\bibfnamefont {S.}~\bibnamefont {P\'eru}},\ }\href {\doibase
  10.1103/PhysRevLett.102.242501} {\bibfield  {journal} {\bibinfo  {journal}
  {Phys. Rev. Lett.}\ }\textbf {\bibinfo {volume} {102}},\ \bibinfo {pages}
  {242501} (\bibinfo {year} {2009})}\BibitemShut {NoStop}%
\bibitem [{\citenamefont {Berger}\ \emph {et~al.}(1984)\citenamefont {Berger},
  \citenamefont {Girod},\ and\ \citenamefont {Gogny}}]{D1S}%
  \BibitemOpen
  \bibfield  {author} {\bibinfo {author} {\bibfnamefont {J.~F.}\ \bibnamefont
  {Berger}}, \bibinfo {author} {\bibfnamefont {M.}~\bibnamefont {Girod}}, \
  and\ \bibinfo {author} {\bibfnamefont {D.}~\bibnamefont {Gogny}},\ }\href
  {\doibase 10.1016/0375-9474(84)90240-9} {\bibfield  {journal} {\bibinfo
  {journal} {Nucl. Phys. A}\ }\textbf {\bibinfo {volume} {428}},\ \bibinfo
  {pages} {23 } (\bibinfo {year} {1984})}\BibitemShut {NoStop}%
\bibitem [{\citenamefont {Otsuka}\ \emph {et~al.}(1978)\citenamefont {Otsuka},
  \citenamefont {Arima},\ and\ \citenamefont {Iachello}}]{OAI}%
  \BibitemOpen
  \bibfield  {author} {\bibinfo {author} {\bibfnamefont {T.}~\bibnamefont
  {Otsuka}}, \bibinfo {author} {\bibfnamefont {A.}~\bibnamefont {Arima}}, \
  and\ \bibinfo {author} {\bibfnamefont {F.}~\bibnamefont {Iachello}},\ }\href
  {\doibase 10.1016/0375-9474(78)90532-8} {\bibfield  {journal} {\bibinfo
  {journal} {Nucl. Phys. A}\ }\textbf {\bibinfo {volume} {309}},\ \bibinfo
  {pages} {1} (\bibinfo {year} {1978})}\BibitemShut {NoStop}%
\bibitem [{\citenamefont {Nomura}\ \emph {et~al.}(2008)\citenamefont {Nomura},
  \citenamefont {Shimizu},\ and\ \citenamefont {Otsuka}}]{nomura2008}%
  \BibitemOpen
  \bibfield  {author} {\bibinfo {author} {\bibfnamefont {K.}~\bibnamefont
  {Nomura}}, \bibinfo {author} {\bibfnamefont {N.}~\bibnamefont {Shimizu}}, \
  and\ \bibinfo {author} {\bibfnamefont {T.}~\bibnamefont {Otsuka}},\ }\href
  {\doibase 10.1103/PhysRevLett.101.142501} {\bibfield  {journal} {\bibinfo
  {journal} {Phys. Rev. Lett.}\ }\textbf {\bibinfo {volume} {101}},\ \bibinfo
  {pages} {142501} (\bibinfo {year} {2008})}\BibitemShut {NoStop}%
\bibitem [{\citenamefont {Nomura}\ \emph {et~al.}(2010)\citenamefont {Nomura},
  \citenamefont {Shimizu},\ and\ \citenamefont {Otsuka}}]{nomura2010}%
  \BibitemOpen
  \bibfield  {author} {\bibinfo {author} {\bibfnamefont {K.}~\bibnamefont
  {Nomura}}, \bibinfo {author} {\bibfnamefont {N.}~\bibnamefont {Shimizu}}, \
  and\ \bibinfo {author} {\bibfnamefont {T.}~\bibnamefont {Otsuka}},\ }\href
  {\doibase 10.1103/PhysRevC.81.044307} {\bibfield  {journal} {\bibinfo
  {journal} {Phys. Rev. C}\ }\textbf {\bibinfo {volume} {81}},\ \bibinfo
  {pages} {044307} (\bibinfo {year} {2010})}\BibitemShut {NoStop}%
\bibitem [{\citenamefont {Nomura}\ \emph {et~al.}(2011)\citenamefont {Nomura},
  \citenamefont {Otsuka}, \citenamefont {Shimizu},\ and\ \citenamefont
  {Guo}}]{nomura2011rot}%
  \BibitemOpen
  \bibfield  {author} {\bibinfo {author} {\bibfnamefont {K.}~\bibnamefont
  {Nomura}}, \bibinfo {author} {\bibfnamefont {T.}~\bibnamefont {Otsuka}},
  \bibinfo {author} {\bibfnamefont {N.}~\bibnamefont {Shimizu}}, \ and\
  \bibinfo {author} {\bibfnamefont {L.}~\bibnamefont {Guo}},\ }\href {\doibase
  10.1103/PhysRevC.83.041302} {\bibfield  {journal} {\bibinfo  {journal} {Phys.
  Rev. C}\ }\textbf {\bibinfo {volume} {83}},\ \bibinfo {pages} {041302}
  (\bibinfo {year} {2011})}\BibitemShut {NoStop}%
\bibitem [{\citenamefont {Ginocchio}\ and\ \citenamefont
  {Kirson}(1980)}]{ginocchio1980}%
  \BibitemOpen
  \bibfield  {author} {\bibinfo {author} {\bibfnamefont {J.~N.}\ \bibnamefont
  {Ginocchio}}\ and\ \bibinfo {author} {\bibfnamefont {M.~W.}\ \bibnamefont
  {Kirson}},\ }\href {\doibase 10.1016/0375-9474(80)90387-5} {\bibfield
  {journal} {\bibinfo  {journal} {Nucl. Phys. A}\ }\textbf {\bibinfo {volume}
  {350}},\ \bibinfo {pages} {31} (\bibinfo {year} {1980})}\BibitemShut
  {NoStop}%
\bibitem [{\citenamefont {Robledo}\ \emph {et~al.}(2008)\citenamefont
  {Robledo}, \citenamefont {Rodr\'{\i}guez-Guzm\'an},\ and\ \citenamefont
  {Sarriguren}}]{robledo2008}%
  \BibitemOpen
  \bibfield  {author} {\bibinfo {author} {\bibfnamefont {L.~M.}\ \bibnamefont
  {Robledo}}, \bibinfo {author} {\bibfnamefont {R.~R.}\ \bibnamefont
  {Rodr\'{\i}guez-Guzm\'an}}, \ and\ \bibinfo {author} {\bibfnamefont
  {P.}~\bibnamefont {Sarriguren}},\ }\href@noop {} {\bibfield  {journal}
  {\bibinfo  {journal} {Phys. Rev. C}\ }\textbf {\bibinfo {volume} {78}},\
  \bibinfo {pages} {034314} (\bibinfo {year} {2008})}\BibitemShut {NoStop}%
\bibitem [{\citenamefont {Rodr\'iguez-Guzm\'an}\ \emph
  {et~al.}(2010{\natexlab{d}})\citenamefont {Rodr\'iguez-Guzm\'an},
  \citenamefont {Sarriguren}, \citenamefont {Robledo},\ and\ \citenamefont
  {Garc\'ia-Ramos}}]{rayner2010pt}%
  \BibitemOpen
  \bibfield  {author} {\bibinfo {author} {\bibfnamefont {R.}~\bibnamefont
  {Rodr\'iguez-Guzm\'an}}, \bibinfo {author} {\bibfnamefont {P.}~\bibnamefont
  {Sarriguren}}, \bibinfo {author} {\bibfnamefont {L.~M.}\ \bibnamefont
  {Robledo}}, \ and\ \bibinfo {author} {\bibfnamefont {J.~E.}\ \bibnamefont
  {Garc\'ia-Ramos}},\ }\href {\doibase 10.1103/PhysRevC.81.024310} {\bibfield
  {journal} {\bibinfo  {journal} {Phys. Rev. C}\ }\textbf {\bibinfo {volume}
  {81}},\ \bibinfo {pages} {024310} (\bibinfo {year}
  {2010}{\natexlab{d}})}\BibitemShut {NoStop}%
\bibitem [{\citenamefont {Otsuka}\ and\ \citenamefont {Yoshida}(1985)}]{PBOS}%
  \BibitemOpen
  \bibfield  {author} {\bibinfo {author} {\bibfnamefont {T.}~\bibnamefont
  {Otsuka}}\ and\ \bibinfo {author} {\bibfnamefont {N.}~\bibnamefont
  {Yoshida}},\ }\href@noop {} {} (\bibinfo {year} {1985}),\ \bibinfo {note}
  {{}JAERI-M (Japan At. Ener. Res. Inst.) Report No. 85}\BibitemShut {NoStop}%
\bibitem [{\citenamefont {Iachello}\ and\ \citenamefont {Arima}(1987)}]{IBM}%
  \BibitemOpen
  \bibfield  {author} {\bibinfo {author} {\bibfnamefont {F.}~\bibnamefont
  {Iachello}}\ and\ \bibinfo {author} {\bibfnamefont {A.}~\bibnamefont
  {Arima}},\ }\href@noop {} {\emph {\bibinfo {title} {The interacting boson
  model}}}\ (\bibinfo  {publisher} {Cambridge University Press, Cambridge},\
  \bibinfo {year} {1987})\BibitemShut {NoStop}%
\bibitem [{\citenamefont {Scholten}\ and\ \citenamefont
  {Blasi}(1982)}]{scholten1982}%
  \BibitemOpen
  \bibfield  {author} {\bibinfo {author} {\bibfnamefont {O.}~\bibnamefont
  {Scholten}}\ and\ \bibinfo {author} {\bibfnamefont {N.}~\bibnamefont
  {Blasi}},\ }\href@noop {} {\bibfield  {journal} {\bibinfo  {journal} {Nucl.
  Phys. A}\ }\textbf {\bibinfo {volume} {380}},\ \bibinfo {pages} {509}
  (\bibinfo {year} {1982})}\BibitemShut {NoStop}%
\bibitem [{\citenamefont {{Brookhaven National Nuclear Data Center}}()}]{data}%
  \BibitemOpen
  \bibfield  {author} {\bibinfo {author} {\bibnamefont {{Brookhaven National
  Nuclear Data Center}}},\ }\href@noop {} {}\bibinfo {howpublished}
  {{http://www.nndc.bnl.gov}}\BibitemShut {NoStop}%
\bibitem [{\citenamefont {Cejnar}\ \emph {et~al.}(2010)\citenamefont {Cejnar},
  \citenamefont {Jolie},\ and\ \citenamefont {Casten}}]{cejnar2010}%
  \BibitemOpen
  \bibfield  {author} {\bibinfo {author} {\bibfnamefont {P.}~\bibnamefont
  {Cejnar}}, \bibinfo {author} {\bibfnamefont {J.}~\bibnamefont {Jolie}}, \
  and\ \bibinfo {author} {\bibfnamefont {R.~F.}\ \bibnamefont {Casten}},\
  }\href {\doibase 10.1103/RevModPhys.82.2155} {\bibfield  {journal} {\bibinfo
  {journal} {Rev. Mod. Phys.}\ }\textbf {\bibinfo {volume} {82}},\ \bibinfo
  {pages} {2155} (\bibinfo {year} {2010})}\BibitemShut {NoStop}%
\bibitem [{\citenamefont {Iachello}(2001)}]{X5}%
  \BibitemOpen
  \bibfield  {author} {\bibinfo {author} {\bibfnamefont {F.}~\bibnamefont
  {Iachello}},\ }\href {\doibase 10.1103/PhysRevLett.87.052502} {\bibfield
  {journal} {\bibinfo  {journal} {Phys. Rev. Lett.}\ }\textbf {\bibinfo
  {volume} {87}},\ \bibinfo {pages} {052502} (\bibinfo {year}
  {2001})}\BibitemShut {NoStop}%
\bibitem [{\citenamefont {Casten}\ and\ \citenamefont
  {Zamfir}(2001)}]{casten2001}%
  \BibitemOpen
  \bibfield  {author} {\bibinfo {author} {\bibfnamefont {R.~F.}\ \bibnamefont
  {Casten}}\ and\ \bibinfo {author} {\bibfnamefont {N.~V.}\ \bibnamefont
  {Zamfir}},\ }\href@noop {} {\bibfield  {journal} {\bibinfo  {journal} {Phys.
  Rev. Lett.}\ }\textbf {\bibinfo {volume} {87}},\ \bibinfo {pages} {052503}
  (\bibinfo {year} {2001})}\BibitemShut {NoStop}%
\bibitem [{\citenamefont {Stone}(2005)}]{stone2005}%
  \BibitemOpen
  \bibfield  {author} {\bibinfo {author} {\bibfnamefont {N.}~\bibnamefont
  {Stone}},\ }\href@noop {} {\bibfield  {journal} {\bibinfo  {journal} {At.
  Data Nucl. Data Tables}\ }\textbf {\bibinfo {volume} {90}},\ \bibinfo {pages}
  {75} (\bibinfo {year} {2005})}\BibitemShut {NoStop}%
\end{thebibliography}%

\end{document}